\documentclass[useAMS,usenatbib,usefloat]{mnras}
\pdfoutput=1
\usepackage{float}
\usepackage{amsmath}
\usepackage{amssymb}
\usepackage{rotating}
\usepackage{tabularx}
\usepackage{gensymb}

\include{journaldefs}
\usepackage[font=footnotesize,labelfont=bf]{caption}
\newcommand{\vsini} {$v$\,sin\,$i$}

\newcommand{\Teff} {$T_{\rm eff}$}
\newcommand{\grav} {log\,{\em g}}

\newcommand{\logq}{log\,{\em Q}~}

\newcommand{\kms}{$\rm km\,s^{-1}$}

\begin{document}

   \title[Analysis of spectra of hot post-AGB candidates]{\bf Analysis of absorption lines in the high resolution spectra of five hot post-AGB candidates}

\author[Herrero et al.]{
{\parbox{\textwidth}{A. Herrero$^{1,2}$\thanks{E-mail: ahd@iac.es}, M. Parthasarathy,$^{3,4}$\thanks{E-mail: m-partha@hotmail.com}, S. Sim\'on-D\'iaz$^{1,2}$, S. Hubrig$^5$,\\G. Sarkar$^{6}$, S. Muneer$^{3}$}
}\\ \\
$^1$Instituto de Astrofis\'ica de Canarias, 38200, La Laguna, Tenerife, Spain\\
$^2$Departamento de Astrof\'isica, Universidad de La Laguna, 38205, La Laguna, Tenerife, Spain\\
$^3$Indian Institute of Astrophysics, Koramangala, Bangalore 560034, India\\
$^4$National Astronomical Observatory of Japan (NAOJ), 2-21-1 Osawa, Mitaka,, Tokyo 181-8588, Japan\\
$^5$ Leibniz-Institute for Astrophysik, Potsdam (AIP), An der Sternwarte 16, D-14482 Potsdam, Germany\\
$^6$ Indian Institute of Technology, Kanpur, India\\
}
   \date{Received 2016; accepted 2016}

  \maketitle
   
\begin{abstract}

  From an analysis of absorption lines in the high resolution spectra we have derived the radial velocities, stellar parameters ($T_{\rm eff}$, \grav, wind-strength parameter \logq and projected rotational velocity) and abundances (C, N, O, and Si ) of IRAS 17460-3114, IRAS 18131-3008, IRAS 19336-0400, LSE 45 and LSE 163. Abundances are found to be solar, except for a low Si abundance in IRAS 19336-0400 and a mild CNO pattern in LSE 163, that rotates at an unusual large rotational velocity for its spectral classification. Combining the stellar parameters information with Gaia DR2 data we are able to derive absolute magnitudes, radii and luminosities and clarify the possible post-AGB nature of the objects. IRAS 17460-3114 and IRAS 18131-3008 are found to be massive OB stars, whereas IRAS 19336-0400 is found to be a post-AGB star, already showing nebular lines in the spectrum. However, we could not confirm the nature of LSE 45 and LSE 163 as post-AGB stars, although their parameters are much more inconsistent with those of massive stars. In both cases, we find a discrepancy between the spectroscopic mass and that derived from the predictions of post-AGB evolutionary tracks. In addition, LSE 45 lacks nebular lines, that are present in IRAS 19336-0400 at a similar temperature. In the case of LSE 163 the rotational velocity (259$\pm$15 \kms)  would be extremely large for a star evolving to CSPN. The combination of this rotational velocity, the high Galactic latitude, slightly large radial velocity and mild CNO enhancement suggests a history of binary interaction.

\end{abstract}


%
%________________________________________________________________

\section{Introduction}

	The final phases of the evolution of low and intermediate mass stars are dominated by the ejection of the matter that will form the Planetary Nebula (PN). This PN becomes optically visible when its progenitor star reaches a temperature high enough to ionize its surroundings. Central stars of planetary nebulae (CSPN) constitute thus a fundamental phase to understand the evolution of the star and its neighbourhood.
		
From an analysis of IRAS data of several hot post-AGB candidates far-IR (IRAS) colours  similar to that of planetary nebulae (PNe) were  detected \citep{partha93a,partha93b,partha00}. A small fraction of massive OB stars and Luminous Blue Variables (LBVs)  with circumstellar dust shells have far-IR (IRAS) colours  similar to that of hot post-AGB candidates detected from the IRAS data. Moreover, some of the high galactic latitude OB stars are in a post-AGB stage of evolution and they do not have circumstellar dust shells \citep{McCausland92, conlon91}. High resolution spectroscopy, chemical composition analysis and multi-wavelength  studies are needed in order to distinguish hot (OB) post-AGB stars from normal OB stars  with circumstellar dust shells, particularly at overlapping luminosities. Spectroscopic studies of hot post-AGB candidates are very important as some of them are rapidly evolving into young low excitation PNe \citep{partha93c, partha95}, but very few hot post-AGB candidates are well studied. 

In order to further understand these transition objects and their differences to more massive OB counterparts we initiated a program to study the selected hot post-AGB candidates from the lists of \cite{partha95} and \citet{partha00}. With this aim in mind we have obtained high resolution spectra of several hot post-AGB candidates with ESO FEROS facility. In this project we have also included selected high galactic latitude line emission OB stars from the paper of \cite{drilling95} \citep[see][]{gauba03}. 

In this paper we present an analysis of the absorption lines in the ESO FEROS high resolution spectra of the hottest targets in our program, namely IRAS 17460-3114, IRAS 18131-3008, IRAS 19336-0400, LSE 45,and LSE 163. We have derived their effective temperature \Teff, logarithmic gravity \grav, wind strength parameter \logq\footnote{the wind-strength parameter, Q, defined as Q=  $\dot{M}/(R v_\infty)^{3/2}$, basically determines the emission in wind recombination lines like H$_\alpha$ \citep[see][]{puls96}}, projected rotational velocity, single-epoch radial velocity and C,N,O,and Si abundances.

\section{Observations}

	The spectra presented in this work were obtained during an observing run in April 14-17, 2006 with the FEROS spectrograph \citep{kaufer99} attached to the MPG/ESO 2.2-m telescope (Proposal ID.77.D-0478A, PI: M.\,Parthasarathy). The resolving power was R= 48\,000, covering from 3700 to 9200 \AA. The log of the observations is given in Tab.~\ref{obstab}. 
Raw data were reduced following standard procedures of bias substraction, flat field correction and wavelength calibration. Finally, we have normalized the spectra by fitting splines to selected continuum points.
	
		In Tab.~\ref{obstab} we give the Signal-to-Noise ratio (S/N) per pixel measured for the spectra presented in this work. It has been measured in the 4720 to 4750 \AA~ region (with small variations for each star). The S/N ratio is usually high, except for IRAS 19336-0400. Although the value for this star may look small for a quantitative analysis, we note that the value given is per pixel, and that the spectra are oversampled. 

   \begin{table*}
      \caption[]{Log of the observations with star identification, FK5 and Galactic coordinates from Simbad, mean observing date and exposure time in seconds. Signal to noise ratios as explained in text. }
         \label{obstab}
         \begin{tabular}{llcccccccccc}
            \hline
            Star      &  Other ID & $\alpha$(J2000) & $\delta$(J2000) & {\it l} (J2000) & {\it b} (J2000) & MJD Observing date & texp & S/N \\
                         &                &                           &                           &                       &                        &  (+2453840)              &  (s)   & \\
            \hline
            IRAS 17460-3114 & HD 161853 & 17:49:16.559 & -31:15:18.06 & 358.4247858 & -01.876644 & 0.401790 & 1200 & 470 \\
            IRAS 18131-3008 &HD 167402 & 18:16:18.688 & -30:07:29.61 &      2.2643758 & -06.392035 & 2.176526 & 2400 & 230 \\
            LSE 45                  &CD-49 8217 & 13:49:17.581 & -50:22:45.54 & 312.2842599 & +11.438876 & 2.120823 & 2100 & 160 \\
            IRAS 19336-0400 & SS441         & 19:36:17.52   & -03:53:25.3   &   34.5758051 & -11.748158 &  3.295044 & 2700  & 50 \\
            LSE 163                & CD-42 8141 & 13:08:46.355 & -43:27:51.22 & 306.2631871 & +19.295353 & 2.084039 & 3000 & 240 \\
            \hline
            \end{tabular}
  \end{table*}

In addition to the observations presented here, there are IUE spectra for IRAS 17460-3114 and IUE and FUSE spectra for IRAS 18131-3008. The quality of these observations is limited, and we will restrict ourselves mainly to the optical analysis, although in Sect~\ref{paramindiv} we comment briefly on them.

\section{Spectral analysis}
\label{param}

\subsection{Line broadening}
	After the data reduction, spectra were corrected from radial velocity. Table~\ref{stardata} gives the B$_T$ and V$_T$ magnitudes from Tycho-2 catalog \citep{hog00} 
for IRAS 18131-3008 and LSE 163. For IRAS 17460-3114, LSE 45 and IRAS 19336-0400 we adopt the values given by the American Association of Variable Stars Observers (AAVSO) in APASS (AAVSO Photometric All-Sky Survey\footnote{Access through https://www.aavso.org/apass. This research was made possible through the use of the AAVSO Photometric All-Sky Survey (APASS), funded by the Robert Martin Ayers Sciences Fund and NSF AST-1412587}) Data Release 10 because they are in better better agreement with Gaia DR2 colors than the Tycho-2 values after the corresponding transformations\footnote{see Gaia Data Release 2, Sect. 5.3.7 in Documentation Release 1.2: https://gea.esac.esa.int/archive/documentation/GDR2/}. In Table~\ref{stardata} we also give the radial velocity, together with the spectral types that were obtained from the literature. Note that these are single-epoch radial velocities. IRAS 17460-3114 for example is a binary and  radial velocity variable \citep{gamen15}.
	
\begin{figure}
   \centering
   \includegraphics[bb= 150 420 445 420, scale=0.30, angle=270]{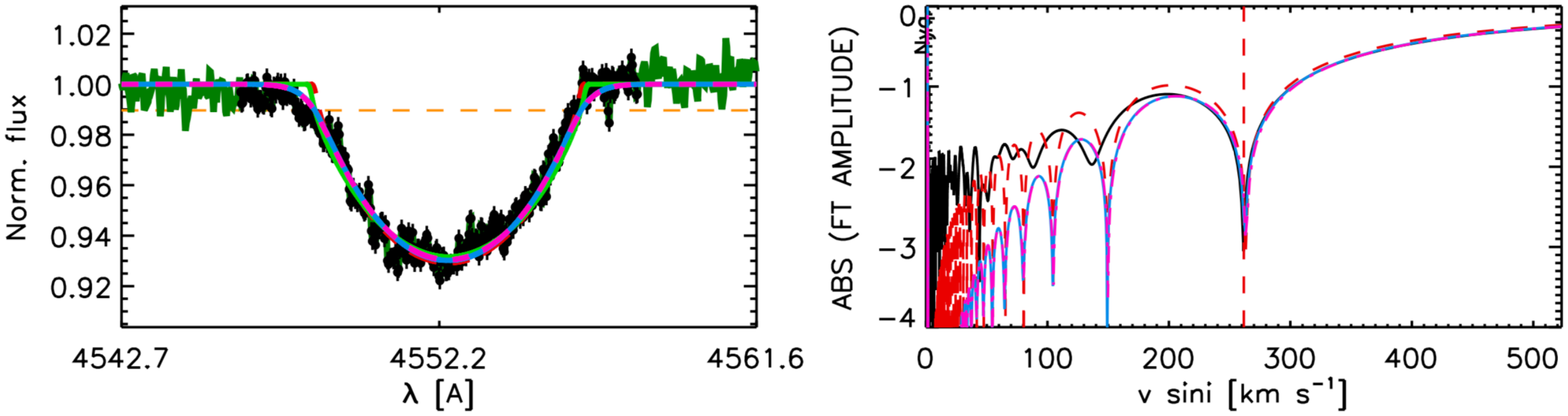}
      \caption{Determination of the broadening (rotation and macroturbulence) in LSE 163. Right: Fourier transform of the observed profile (solid black line) and of the theoretical rotational profile with \vsini= 259 \kms. The central vertical line shows the position of the first zero, and the one close to the ordinate axis, the Nyquist frequency. Left: Fits using the Goodness of Fit (GOF) method. Note the pure rotational profiles departing at the left wing. See text for more details.}
         \label{vrot163}
   \end{figure}

	We then determined the rotational and {\it macroturbulent}\footnote{We follow here the usage in calling macroturbulence the extra broadening in the spectral lines of yet unknown origin. It is basically what remains after taking natural, collisional, thermal, microturbulence and rotation into account} velocities, following the procedures described in \cite{SH14}. Briefly, the rotational velocity is determined in two ways.  In the first method, we carry out a goodness-of-fit (GOF), comparing the observed profile with a profile convolved with rotation and a radial-tangential macrotubulent profile.  In the second method, we use the Fourier transform of the spectral line, assuming that the first zero of the transform is due to the rotational profile (although sometimes, with very low \vsini, {\it micro}turbulence may introduce a zero before those due to rotation). Macroturbulence is obtained then by applying again the goodness-of-fit method, but now with the vsini value obtained from the Fourier transform. Ideally, both methods should give the same result. When this does not happen, we make use of extra information, like visual inspection of the profiles to detect blends, assymetries or other conditions that may hinder the analysis. 
	Strong unstaurated and unblended metal lines are the preferred profiles for these determinations. In OB stars, the best profiles in the optical domain for the determination of rotational and macroturbulent velocities are \ion{Si}{iii} 4552 and \ion{O}{iii} 5592 \AA~lines. The lines will not be equally strong in our objects, with the \ion{O}{iii} increasing towards hotter objects and the \ion{Si}{iii} towards cooler ones in our stellar sample. After inspecting the spectra, we have used \ion{O}{iii} 5592 \AA~ for IRAS 17460-3114 (moreover, the line will be weak in the secondary), both lines for IRAS 18131-3008 and LSE 45, and the \ion{Si}{iii} 4552 \AA line for IRAS 19336-0400 and LSE 163. Both lines give very similar results when they are of the same quality in the spectra. Table~\ref{stardata} gives the values obtained for our objects. The rotational velocities of IRAS 17460-3114 (164$\pm$9 km s$^{-1}$) and especially LSE 163 (259$\pm$15 km s$^{-1}$) are larger than typical for CSPN (while massive OB stars can rotate at any velocity between a few \kms and up to 600 \kms -- see \cite{ramirezag13, SH14, dufton11}--, CSPN rotate at less than 100 \kms; \citep[see f.e.][who obtain values below 100 \kms, and mostly below 50 \kms]{mello12, prinja12}. We use LSE 163 to illustrate the method in Fig.~\ref{vrot163}. The left plot shows the fits to the observed spectrum using the Goodness of Fit method. The blue fit corresponds to a rotational plus radial-tangential macroturbulent profile (the chosen one) and the green one to a pure rotational profile, whereas the  purple and red fits correspond to the same profiles but fixing the rotational velocity to the value obtained with the Fourier method (right plot). We see that the the only difference is that pure rotational profiles fail in the wings, indicating the need of an extra broadening mechanism, but fit the line core. All methods give rotational velocities between 259 and 272 km s$^{-1}$. Remarkably, the GOF and Fourier methods give the same projected rotational velocity of 259$\pm$15 km s$^{-1}$, as can be seen in the right plot, where the blue (rotational plus radial-tangential macroturbulent profile) and red (pure rotational, fitting the first zero in the Fourier transform) are shown. More details about the method and error determination can be found in \cite{SH14}.

   \begin{table*}
      \caption[]{Spectral types, photometric magnitudes (as explained in text)  and radial, rotational and macroturbulent velocities (all in km s$^{-1}$). Note that IRAS 17460-3114 is a radial velocity variable \citep{gamen15} Spectral classification references: {\tiny (1) \citet{gamen15} ; (2) \cite{garrison77};  (3) \citet{gauba03}; (4) \citet{partha00}} }
         \label{stardata}
         \begin{tabular}{llccrrrc}
            \hline
            Star      & SpT & B  &  V  & $V_{\rm rad}$ & $Vsini$ & $V_{\rm mac}$ & Ref \\
            \hline
            IRAS 17460-3114 & O8 V(n)z + B1-3 V & 8.094$\pm$0.029 & 7.853$\pm$0.045 & -65.4 & 164$\pm$9 & 112$\pm$ 12 & 1 \\
            IRAS 18131-3008 & B0 Ib      &  8.939$\pm$0.019 & 9.025$\pm$0.020 & 23.4 &   44$\pm$3 &  81$\pm$9 & 2\\
            LSE 45                 & B2 I         & 10.808$\pm$0.027& 10.814$\pm$0.093 & 77.4 &   21$\pm$3 & 49$\pm$5 & 3 \\
            IRAS 19336-0400& B1 Iaep   & 13.478$\pm$0.126& 13.096$\pm$0.111  & 67.4 &   84$\pm$9 &   4$\pm$6 & 4\\
            LSE 163               & B2 I         & 10.356$\pm$0.027 & 10.457$\pm$0.038 & -65.1 &  259$\pm$15 &104$\pm$12 & 3 \\
            \hline
            \end{tabular}
  \end{table*}

  \subsection{Stellar parameters}
  The stellar and wind parameters of our stars were determined  using the IACOB Grid-Based Automatic Tool
\citep[{\sc iacob-gbat}, ][]{ssimon11, holgado18}. The tool is based on the standard criteria developed for the quantitative spectroscopic analysis of O stars 
\citep[see, e.g.][]{h92,h02,repolust04}. It applies a $\chi^2$ algorithm to a large pre-computed 
grid of {\sc FASTWIND}
synthetic spectra. {\sc FASTWIND} \citep{santolaya97, puls05} is a model atmosphere code with spherical geometry, Non-Local Thermodynamic Equilibrium, mass-loss and line blanketing. The elements considered in this work (H, He, C, N, O and Si ) are treated explicitly through detailed model atoms, whereas the rest of relevant elements (including for example Fe and Ni) are treated with different approximations. Details can be found in \cite{puls05}. The grid\footnote{The grid was computed using the Condor workload management system (http://www.cs.wisc.edu/condor/) implemented at the Instituto de Astrof\'isica de Canarias.} comprises $\sim$\,180000 atmosphere models, defined by \Teff, \grav, \logq (with no clumping), YHe (the helium abundance with respect to Hydrogen, by number), the microturbulent velocity $\xi$ and the exponent of the wind velocity law\footnote{we adopt a wind velocity law of the form $v(r)= v_\infty (1-\frac{R_*}{r})^\beta$, where $v_\infty$ is the wind terminal velocity and $R_*$ is the stellar radius}, $\beta$, covering a wide 
range of stellar and wind parameters given in Tab.~\ref{grid} and optimized for the analysis of OB-stars. The model spectra are compared to the observed ones, and for each model $\chi^2$ is calculated. The tool delivers the best stellar parameters together with their $\chi^{2}$ distributions, and assigns them formal errors. We found most of these uncertainties to be quite low. Following the reasoning in \cite{sabin14}, who argue that these formal errors do not take other uncertainties like continuum normalization into account, we have assigned minimum errors of 0.1 dex to \grav, and 0.02 to YHe.
Results for the five stars in our sample are presented in Tab.~\ref{stparam} and the line fits are presented in Figs.~\ref{fits1},~\ref{fits2},\ref{fits3}-\ref{fits5}. According to \cite{massey13} the gravities obtained with {\sc FASTWIND} are in average lower by 0.12 dex than those obtained with the {\sc CMFGEN} code \citep{HM98}. Similar results, although slightly lower, have been more recently obtained by \cite{sabin17} and \cite{holgado18} in the LMC and with Galactic standards, respectively. They estimate that {\sc FASTWIND} gravities may be low by 0.06-0.09 dex, producing uncertainties of 10-20 $\%$ in the spectroscopic masses, which is within our error bars.

\begin{table}
  \caption[]{Parameter range for the grid of {\sc FASTWIND} model atmospheres}
     \label{grid}
     \begin{tabular}{ll}
     \hline
     Parameter & Range \\
     \hline
     \Teff (kK) & 22.0-55.0 with $\Delta$\Teff= 1.0 \\
     \grav (dex) & 2.60-4.20 with $\Delta$\grav= 0.1 \\
     \logq (dex) & $-$11.7 to $-$12.7 with $\Delta$\logq= 0.2 \\
              & plus $-$13.0, $-$13.5, $-$14.0, $-$15.0 \\
     YHe & 0.06, 0.10, 0.15, 0.20, 0.25, 0.30 \\
     $\xi$ (km s$^{-1}$) & 5-20 with $\Delta\xi$= 5 \\
     $\beta$ & 0.8, 1.0, 1.2, 1.5, 1.8 \\
     \hline
     \end{tabular}
\end{table}

   \begin{table*}
      \caption[]{Main spectroscopic parameters and stellar abundances (see text). For IRAS 17460-3114 we quote the values of the combined spectrum with increased uncertainties, except for the helium abundance, that is adopted as solar. Solar abundances are taken from \cite{asplund09}. Note that the gravity of LSE 163 has to be corrected from centrifugal acceleration, resulting in a total \grav= 3.52.}
         \label{stparam}
         \begin{tabular}{llccccccccc}
            \hline
            Star      & \Teff & \grav & YHe & \logq &      C         &      N       &      O       &      Si                      \\
            \hline
            IRAS 17460-3114 & 37.0$\pm$0.6 & 4.13$\pm$0.12 & {\it 0.10} & $-$13.3$\pm$ 0.4 &                          & 7.85$\pm$0.14 &                &                 \\
            IRAS 18131-3008 & 29.8$\pm$0.4 & 3.30$\pm$0.10 & 0.10$\pm$0.02 & $-$13.3$\pm$0.3 & 8.35$\pm$0.26 & 7.81$\pm$0.19 & 8.71$\pm$0.24 & 7.54$\pm$0.13 \\
            LSE 45                 &  29.0$\pm$0.2 & 3.52$\pm$0.10 & 0.10$\pm$0.02 & $-$13.5$\pm$0.15 & 8.37$\pm$0.11 & 7.65$\pm$0.18 & 8.68$\pm$0.17 & 7.49$\pm$0.12  \\
            IRAS 19336-0400&  27.7$\pm$1.1 & 3.36$\pm$0.15 & 0.10$\pm$0.06 &  $-$12.4$^{+1.5}_{-0.10}$ & 8.55$\pm$0.18 &     ---                 & 8.80$\pm$0.33 &  6.93$\pm$0.26       \\
            LSE 163               &  23.4$\pm$0.4 & 3.12$\pm$0.10 & 0.10$\pm$0.02 & $< -$13.5 & 8.26$\pm$0.17 & 8.00$\pm$0.15 & 8.59$\pm$0.16 & 7.64$\pm$0.21          \\
            \hline
Solar                     &               &            &        &      & 8.43$\pm$0.05 & 7.83$\pm$0.05 & 8.69$\pm$0.05 & 7.51$\pm$0.03 \\
\hline
\            \end{tabular}
  \end{table*}
  
\subsubsection{Comments on the stellar parameters determination of individual stars}
\label{paramindiv}
\begin{figure}
   \centering
   \includegraphics[bb= 0 0 600 770, scale=0.35,angle=90]{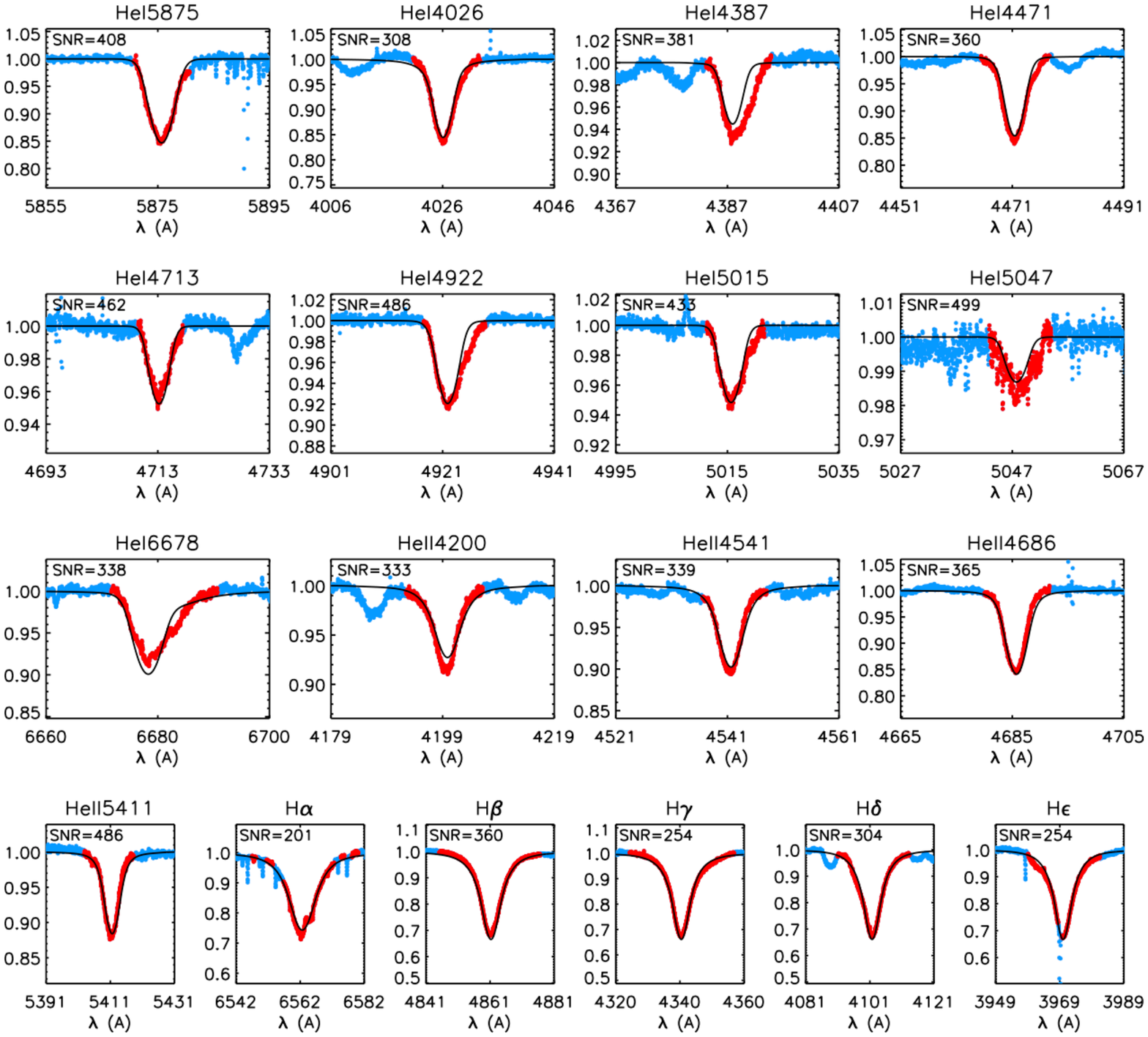}
      \caption{The fit to the H-He spectrum of IRAS 17460-3114 with the parameters indicate in Table~\ref{stparam}. The red points in the observed profiles indicate the data considered for the analysis, whereas blue points were not taken into consideration. Calculated profiles are shown in black with a solid line. Abcissa gives the wavelength in \AA~ and ordinate the relative flux}
         \label{fits1}
   \end{figure}

\begin{figure}
   \centering
   \includegraphics[bb= 0 0 600 770, scale=0.35,angle=90]{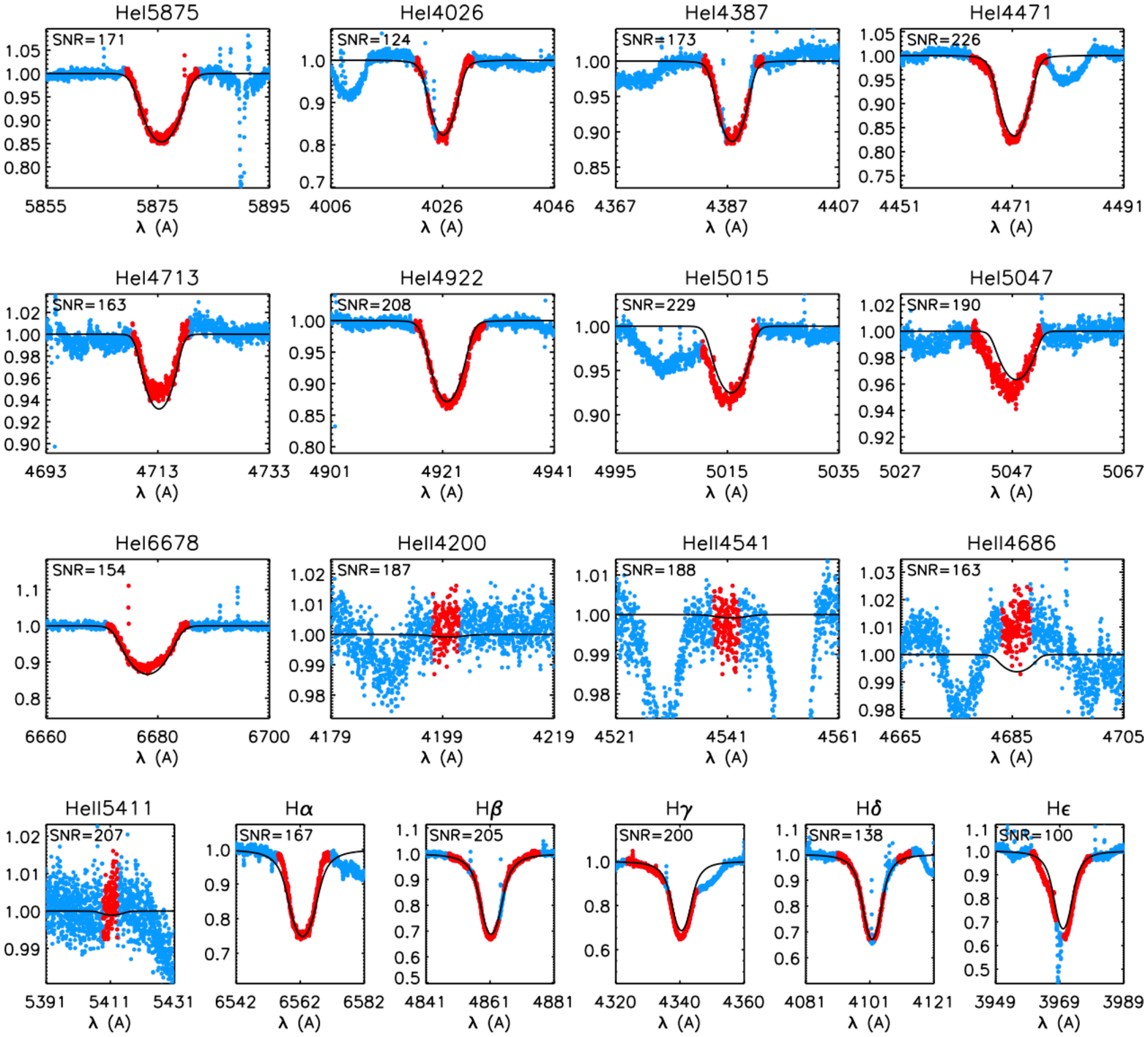}
      \caption{Same as Fig.~\ref{fits1} for LSE 163}
         \label{fits5}
   \end{figure}

{\it IRAS 17460-3114 (HD\,161\,853, SAO 209\,306) } The fits to the lines of this star can be seen in Fig.~\ref{fits1}. They are good except for the \ion{He}{i} lines, that show clear contamination in their red wings, consistently with contamination from a cooler companion as indicated in the GOSSS catalog \citep{maiz16} and in \cite{gamen15}. \ion{He}{ii} lines on the other hand do not show any contamination, nor is this detected in other lines in the spectrum. This includes the Balmer lines, that however could be slightly affected, resulting in a slightly large \grav~ value (which would also alter the \Teff\, value). The derived He abundance is below the solar value value of 0.09$\pm$0.01 \cite{asplund09}, which is again consistent with the contamination of the spectrum (that may produce a line dilution because of the additional continuum). These findings are in agreement with those by \cite{mello12} and \cite{gamen15}, who also found evidences of a companion. Thus we conclude that this star is a binary, with its primary component being an O star and a companion being a fainter B star. To estimate the effect of the companion on the stellar parameters, we have assumed that the secondary contributes between 10$\%$ and 30$\%$ to the continuum flux from the system and have redone the analysis only with the \ion{He}{ii} lines (assuming a solar helium abundance). The changes in the stellar parameters are within uncertainties, except for gravity in the case of a serious (30$\%$) contamination. However, we exclude this case, as it results in Balmer lines that are far from being reproduced by the models (whereas we know that present-day models reproduce fairly well the Balmer spectrum of O8 V stars). Therefore we keep the values from the original analysis, increasing the uncertainties and noting that gravity could be slightly lower than quoted, but always within the given errors. For the Helium abundance we adopt a solar value. Finally, we note that there is an IUE low resolution spectrum in the INES database showing an unsaturated \ion{C}{iv} absorption profile, insufficient to derive an accurate terminal velocity. 

\begin{figure}
   \centering
   \includegraphics[bb= 0 0 600 770, scale=0.35,angle=90]{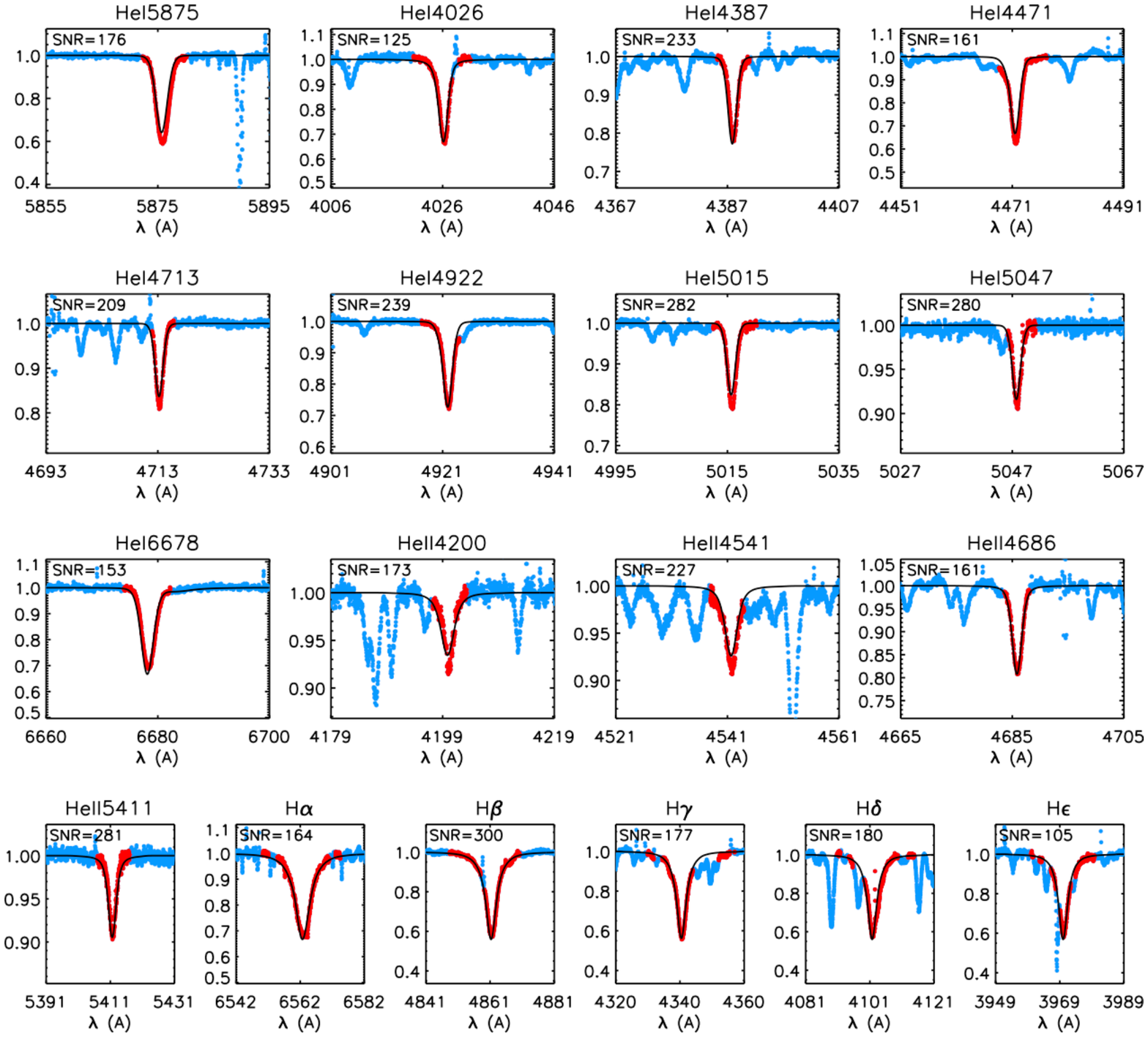}
      \caption{Same as Fig.~\ref{fits1} for  IRAS 18131-3008}
         \label{fits2}
   \end{figure}

{\it IRAS 18131-3008 (HD\,167\,402, LS\,4835)} The derived rotational velocity is quite low, of the order of 40 km s$^{-1}$. \cite{savage01} found a larger \vsini~ from the UV spectrum, which is explained because the broadening is dominated by the macroturbulent profile. The combined UV spectrum from the INES archive does not show a clear \ion{C}{iv} 1550 P-Cygni profile, although it is nearly saturated, indicating a wind terminal velocity about 2\,000 km s$^{-1}$. The global fit to the H-He spectrum is again vey good, as can be seen in Fig.~\ref{fits2}. The He abundance is normal, with a very small uncertainty. 
 
\begin{figure}
   \centering
   \includegraphics[bb= 0 0 600 770, scale=0.35,angle=90]{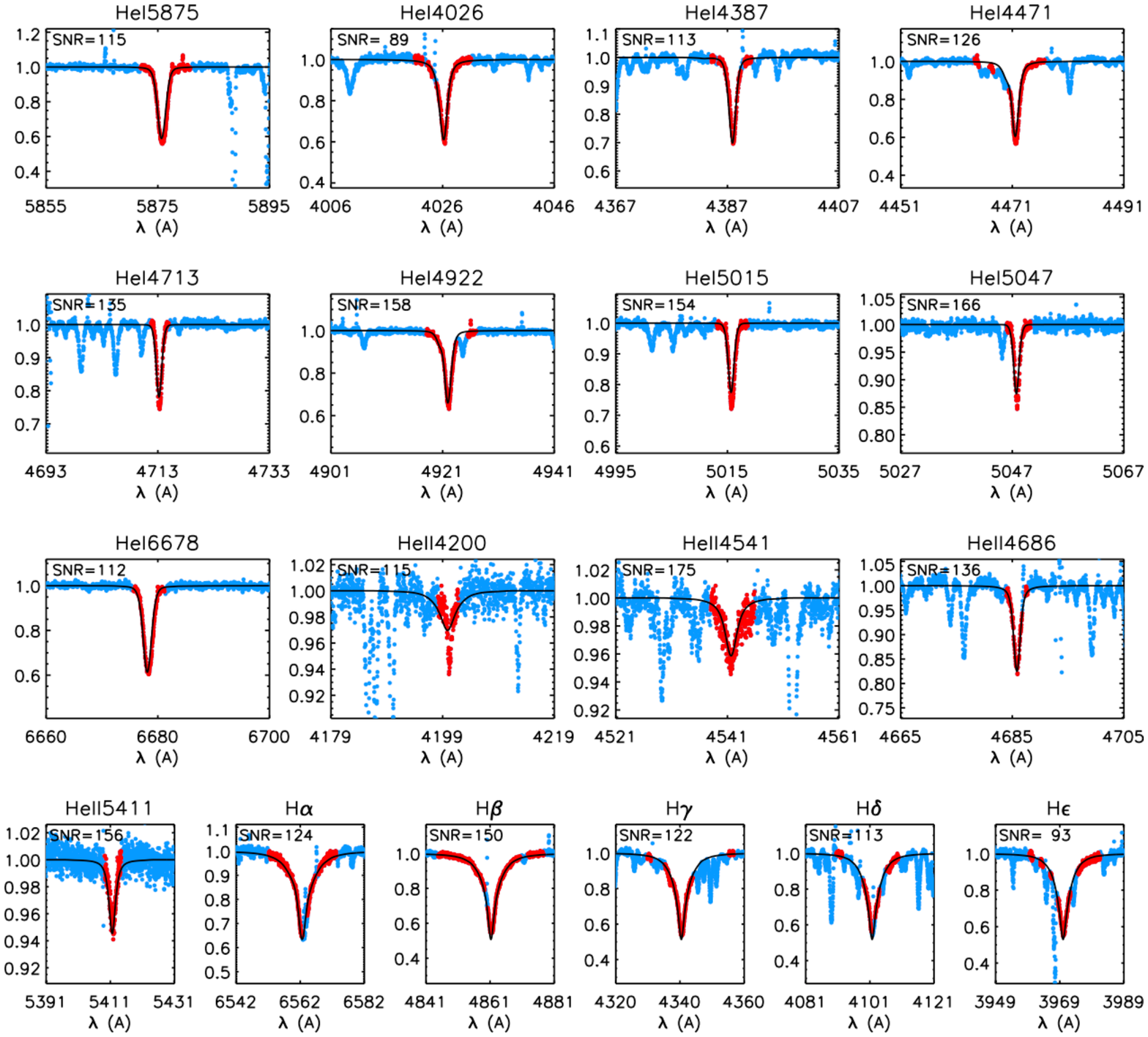}
      \caption{Same as Fig.~\ref{fits1} for LSE 45}
         \label{fits3}
   \end{figure}

 {\it LSE 45 (CD -49 8217)} This star has a large radial velocity, a very low rotational velocity and a modest macroturbulent broadening.  The fit to the H-He spectrum, shown in Fig.~\ref{fits3}, only shows small problems in the cores of some \ion{He}{} lines and in the fit to \ion{He}{ii} 4200. The latter may be produced by the presence of a \ion{N}{iii} line at the same position, and is a characteristic of stars with \Teff$\sim$30\,000 K at very high spectral resolution (at low resolution, this fact may mislead the analysis if too much weight is given to \ion{He}{ii} 4200 \AA). 
 
\begin{figure}
   \centering
   \includegraphics[bb= 0 0 600 770, scale=0.35,angle=90]{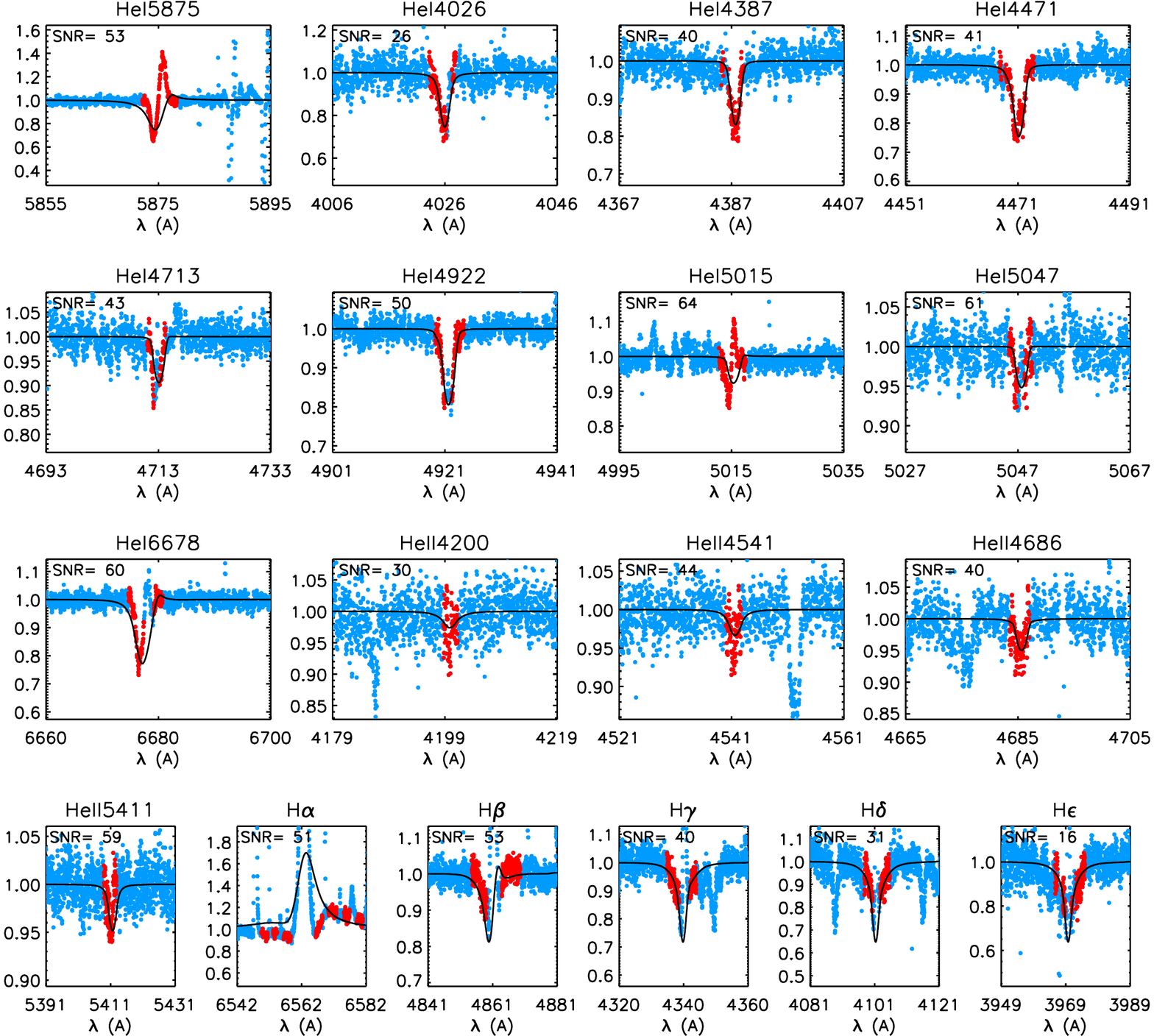}
      \caption{Same as Fig.~\ref{fits1} for IRAS 19336-0400}
         \label{fits4}
   \end{figure}

   {\it IRAS 19336-0400 (G034.5-11.7)} This is the faintest target, which reflects in the low S/N ratio.  The stellar spectrum is contaminated by nebular emission lines. The low S/N produces larger uncertainties for \Teff~ and the Helium abundance than for the other cases. We note that the wind strength parameter, \logq, is the largest of all values in spite of a relatively low temperature. This large value is not the consequence of the contamination by the nebular spectrum as revealed by the P-Cygni profiles in some \ion{He}{i} lines. These P-Cygni profiles are not well reproduce by the final model (see Fig.~\ref{fits4}), because our automatic method favours the fit to the photospheric lines. To account for it we have increased the uncertainty in \logq towards higher values (to this aim, we calculated several models beyond the limits of our grid in Tab.~\ref{grid}, checking that the photospheric parameters are not significantly affected). The H$_\alpha$ profile shows a broad emission in its red wing that has a clear stellar origin. The Helium abundance is solar.
    
   {\it LSE 163 (CD -42 8141)} This is the star with the lowest temperature in our sample, and also with the largest rotational velocity (259$\pm$15 km s$^{-1}$), implying a significant correction to gravity because of centrifugal acceleration \citep[see][]{h92}. This is much larger than typical values for CSPN (\cite{mello12, prinja12} give 2-85 \kms). The broad lines indicating a large rotational velocity are clearly seen in the spectrum, and in Fig.~\ref{vrot163} we see the quality of the fits used for its determination. The derived Helium abundance is normal, but compatible with a mild contamination with CNO processed material at the surface (as found in the next section). 
   
\section{Abundance determination}
\label{abund}

To determine abundances of C, N, O and Si we have first measured their equivalent widths (EWs) following the methods described in \cite{ssimon10}. Briefly, we set up a line list containing all relevant lines. A wavelength range around the line center is defined according to the rotational velocity, to constrain the line limits. EWs are then measured automatically, both fitting a Gaussian and integrating the area below the continuum in the delimited region. Most of the times both measurements are consistent within errors. By visual inspection we discard those measurements that present obvious problems, like a clear overlap with adjacent lines or poor fit to a Gaussian. Uncertainties are estimated by varying the continuum level according to the S/N ratio. This way we obtain the values that are listed in Tables.~\ref{tabewsc} to \ref{tabewssi}.
 
 Once we have measured the EWs of each element we have determined their abundances following the procedures presented in \cite{ssimon10}. We begin with the analysis of Si, determining the value of microturbulence that gives a zero slope in the Abundance versus Equivalent Width diagram. When more than one ionization stage is present in the spectrum, we correct the effective temperature, if required, to avoid a difference in the abundances indicated by different ionization stages, and determine again the microturbulence. Thus, the effective temperature (and correspondingly, the surface gravity) indicated by the ionization balances of Si and He may be slightly different by 200-300 K \cite[see also][who finds similar results]{ssimon10}. Once we have determined the microturbulence, we can calculate the abundance for each individual metal line, where we exclude those that are not present in our atomic models. The final abundance is determined by averaging all available lines, after having discarded those departing significantly (more than 2$\sigma$). Table~\ref{stparam} gives the derived abundances together with their uncertainties. For elements other than Si we adopt the values of \Teff~and \grav~that agree with the Si ionization equilibrium, because they give a better consistency among the metallic lines, and follow the same procedure. Exceptions are LSE 163 (for which only \ion{Si}{iii} lines are detected because of the large rotational velocity), and the N abundance of IRAS 17460-3114 because of its dependence on the mass-loss rate.
 
For the determination of the uncertainties given in Tab.~\ref{stparam} we have considered the scatter in individual line abundances, the uncertainties in stellar parameters ($\Delta T_{\rm eff} \approx $ 1000 K, $\Delta {\rm log}g \approx $ 0.1 dex) and the additional scatter introduced by the change in effective temperature required by the He and Si ionization balances (this last one is usually dominated by the errors in stellar parameters). Errors have been added quadratically to give the values quoted in Tab.~\ref{stparam}.

\subsection{Notes on individual abundance determinations}

\begin{figure}
   \centering
   \includegraphics[bb=20 20 730 540, scale=0.35,angle=180]{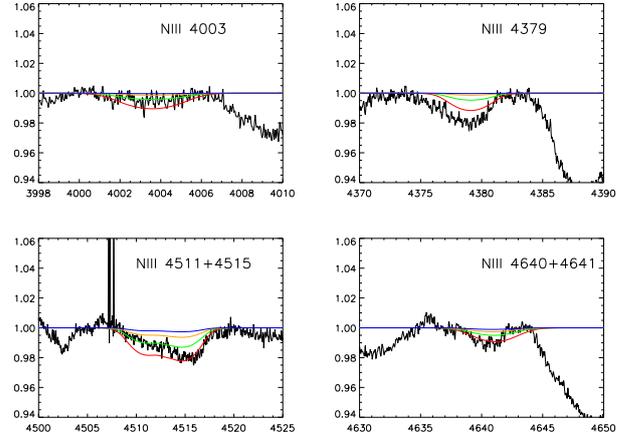}
      \caption{Behaviour of the \ion{N}{iii} lines for different N abundances at the stellar parameters in our grid close to those derived for IRAS 17460-3114. The abcissa is the wavelength in Angstroms and the ordinate gives the relative flux. Models have Teff= 37\,000 K, \grav= 4.00, YHe=0.10 and \logq= -13.5. Colors are for models with different N abundance:  log(N/H)+12= 8.00 (red, deeper line), 7.60 (green), 7.20 (orange) and 6.80 (blue, shallower line). The best fit is obtained for a N abundance between log(N/H)+12= 7.60 and 8.00} 
         \label{nabun}
   \end{figure}

{\it IRAS 17460-3114} This is the hottest star in our sample. Its relatively large rotational and macroturbulent velocities make the determination of equivalent widths less accurate than for cooler and slower rotating stars. As we have a lower number of available lines for our atomic models, the determination of microturbulence is not possible. We adopted a range of microturbulent velocities between 5 and 15 km s$^{-1}$ and directly compared by eye with calculated models. Moreover, because of the high temperatures, only the N model atoms available for FASTWIND are reliable, and we thus restrict ourselves to this element. We have analyzed the behaviour of \ion{N}{iii} lines at 4003, 4379, 4511-15 and 4640 \AA~ around the stellar parameters determined for the star. For an initial microturbulence of 10 km s$^{-1}$ we find that \ion{N}{iii} 4003 and 4511-15 \AA~ give always consistent values around log(N/H)+12= 7.80 without a strong influence of the chosen ($T_{\rm eff}$, log$g$) pair. \ion{N}{iii} 4640 \AA~ shows a larger dependence on the stellar parameters due to its sensitivity to wind strength (and thus to effective gravity). Nevertheless, for the derived stellar parameters we obtain a result consistent with that of \ion{N}{iii} 4003 and 4511-15 \AA, namely log(N/H)+12= 7.90. The line \ion{N}{iii} 4379 \AA~ gives very high values for the N abundance that are in clear disagreement with the other lines, and we discard it. Thus we adopt for the star log(N/H)= 7.85$\pm$0.10, where the uncertainty comes from the line dispersion and the stellar parameters uncertainties. Finally, we have considered the effect of varying the microturbulence between 5 and 15 km s$^{-1}$ and find a dispersion in the results of $\pm$0.1 dex. Thus we finally adopt log(N/H)= 7.85$\pm$0.14 by adding both error sources quadratically. In Figure~\ref{nabun} we show the behaviour of the \ion{N}{iii} lines for different N abundances at the grid stellar parameters closest to the final values.

{\it IRAS 18131-3008} This star has a relatively low rotational velocity and thus the determination of abundances does not present particular problems when following the procedure described at the beginning of this section, except for the C atom, for which the \ion{C}{ii} lines point to a lower than solar value, in constrast to the \ion{C}{iii} lines. The average result, still compatible with the solar value, has thus a relatively large uncertainty. We also get a relatively large error bar for O, but this time it is the result of comparable values for the stellar parameters and line dispersion uncertainties (we are usually dominated by the former one). Lines that were problematic (deviating more than 2$\sigma$ from the average value) have been marked in the Tables in Appendix A. The most significant were the \ion{Si}{iv} and \ion{O}{ii} lines at 4089, which could not be adequately deblended. Thus the abundances derived for all elements agree with the solar abundances given by \cite{asplund09}, with somewhat large uncertainties for C and O.

{\it LSE 45} This is the star with the lowest rotational velocity in our sample, which favours lower uncertainties. The largest difficulties were in the determination of the N abundance because fo the impact of the stellar parameters uncertainty, that dominates the error in the abundance because of the sensitivity of the N lines to changes in the stellar parameters. All N lines gave consistent results, except \ion{N}{ii} 5045 \AA. The \ion{O}{ii} 4089 \AA~ line was this time consistent with the rest of \ion{O}{ii} lines, but its \ion{Si}{iv} blend was problematic, deviating more than 2$\sigma$ from the average, and therefore we decided to skip both lines in the analysis. Again, all abundances are consistent with the solar abundances from \cite{asplund09}. The nominal value obtained for N is slightly low, but as we do not find any hint of anomalies in other elements we consider the chemical composition of the star completely normal.

{\it IRAS 19336-0400} The low S/N ratio rather than the modest rotational velocity of this star is the reason for the large uncertainties and our impossibility to determine the N abundance, as we only detect one line (\ion{N}{iii} 4640 \AA) that is blended with \ion{O}{ii} and \ion{C}{iii}. The values we obtain for C and O are consistent (within the large uncertainties) with the solar values from \cite{asplund09}. However, the Si abundance is sub-solar. This is partly a consequence of the difficulty to derive the microturbulence and trace an accurate normalization of the continuum (particularly around H$_\delta$) at this low S/N. Because of the large uncertainty and the normal abundances derived for C and O we consider that the star has most probably a normal (solar) chemical composition, although we quote our value in Table~\ref{stparam} for future checks.

 {\it LSE 163} The large rotational velocity of this star makes difficult to measure accurately the equivalent widths and dilutes the weakest lines, that are not detected anymore. The low number of available lines renders the determination of the microturbulence very difficult. In this case, we try to reduce the dispersion given by the different spectral lines of each element. We use \ion{O}{ii} 4414-16 \AA. These lines are blended, and we assume that the individual equivalent widths are in the proportion given by their oscillator strengths (as they belong to the same multiplet) to give the total equivalent width measured. They give a good agreement for a microturbulence of 7.0 kms$^{-1}$ and its uncertainty is dominated by that in the stellar parameters. For Si, we use \ion{Si}{iii} 4552-67 \AA~ and a microturbulence slightly larger than for O, as found for other analyzed stars (9.0 km s${-1}$). In this case, the uncertainty is dominated by the dispersion between the lines, reflecting the difficulty of accurate measurements. For N, we exclude the \ion{N}{ii} 4045 \AA~ line, as it is contaminated by \ion{He}{i} and use the \ion{N}{ii} lines at 5005-7 \AA~ (although they are also blended by the nearby \ion{He}{i} 5015 \AA~ line). Adopting the microturbulence of Si, we find log(N/H)= 8.00$\pm$0.15, a value pointing to CNO contamination at the surface. This agrees with the decrease in C abundance implied by the \ion{C}{ii} 4267 \AA~ line (log/C/H)+12= 8.26$\pm$0.17. The \ion{C}{ii} 6578-82 \AA~ lines are strongly blended and near H$_\alpha$. We use the same procedure as for Si, i.e., we measure the total equivalent width and assigne the individual values proportionally to the log(gf) values. Whereas \ion{C}{ii} 6578 \AA~ requires a very high C abundance, probably pointing to some residual contamination by H$_\alpha$, \ion{C}{ii} 6582 \AA~ fully agrees with the abundance derived for \ion{C}{ii} 4267 \AA, that we adopt as final value for this star. With respect to solar values, we obtain [N/H]= +0.17$\pm$0.16 and [C/H]= -0.17$\pm$0.18. Although the uncertainties are of the same order than the effect, the consistency between the results of N and C strongly point to a moderate CNO contamination at the surface.

  \section{Distances, radii, luminosities and masses}
\label{distances}

We present in Tables~\ref{dr2data} and~\ref{dr2datab} the data from the second Gaia Data Release (DR2) \citep{gaia18} that will be used in the discussion together with the stellar parameters obtained in this work. Distances are taken from the work by \cite{bailer18}, who derive them from the Gaia DR2 parallaxes through bayesian inference, adopting adopting as prior an exponentially decreasing space density with a scale length that depends on the Galactic latitude and longitude (for a discussion of the prior impact \citep[see e.g.][]{bailer15}. All our targets fulfill the key quality requirement of at least 8 visibility periods used and a Renormalized Unit Weight Error (RUWE) smaller than 1.40 \citep{lindegren18b}, except IRAS 19336-0400, for which a slightly larger value of RUWE= 1.61 is obtained. Nevertheless, the distances obtained for these stars. Most of our stars are at relatively large distances, so that the systematic offset up to 0.1 mas present in Gaia DR2 parallaxes introduces a large uncertainty for small parallaxes \citep[e.g.][]{lindegren18}. Note that this is not included in the errors given in Tab.~\ref{dr2data}. 

IRAS 17460-3008 has a relatively large DR2 parallax and the distance given by \cite{bailer18} is 1.249$^{+0.141}_{-0.114}$ kpc. However, IRAS 17460-3008 (B-V) color does not agree with the recent extinction map presented by \cite{chen19} at this distance. According to this map, IRAS 17460-3008 should be at a distance of at least 1.5 kpc, incompatible with the Gaia parallax. We note that the \cite{chen19} map averages over 6$\times$6 arcmin and 200 pc and thus a given object in the field may depart from the average behaviour. Being IRAS 17460-3008 a binary we may wonder whether the Gaia parallax is affected by the orbital motion. According to \cite{gamen15} the size of the orbit is too small for any reasonable inclination to produce an effect the on the parallax beyond the quoted uncertainty (note that for any reasonable stellar mass the inclination cannot be significantly lower than 45$^\circ$). Moreover, Gaia colors and the photometry given in Tab.~\ref{stardata} agree with each other (after the corresponding transformations) and indicate an extinction larger than the \cite{chen19} map at the Gaia distance. 
Therefore we keep the Gaia distance and conclude that the extinction difference with the \cite{chen19} map is due to IRAS 17460-3008 departing from the average behaviour in its pixel.

For IRAS 18131-3008 the quoted parallax is very close to zero (and negative), with an uncertainty that is 10 times larger than the listed value. Therefore we see its Gaia distance as unreliable. For the other targets (except IRAS 17460-3114) distances have to be taken also with extreme caution, particularly that of IRAS 19336-0400, with a RUWE value larger than the recommended maximum.

\begin{figure}
   \centering
   \includegraphics[bb=0 0 730 610, scale=0.35,angle=180]{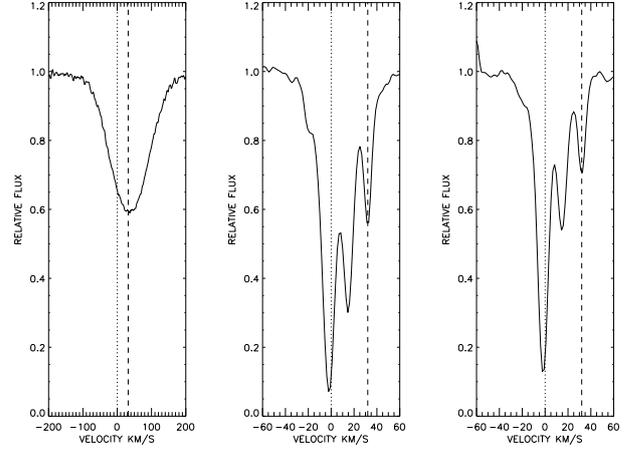}
      \caption{The \ion{He}{i} 5876 (left panel) and \ion{Na}{i} 5890, 5896 (middle and right panels) lines on the IRAS 18131-3008 spectrum in velocity space. The dotted vertical lines mark the zero velocity and the dashed lines mark the velocity of the third absorption component in the Na spectrum. There is a perfect coincidence between the stellar displacement and that of this third component.}
         \label{nai}
   \end{figure}

To get a different estimation for IRAS 18131-3008 we rely on the observed spectrum. The stellar spectrum is redshifted by about +31.5 km s$^{-1}$ with respect to the zero-velocity interstellar \ion{Na}{i} lines at $\lambda\lambda$5890, 5896 \AA.  The \ion{Na}{i} lines display three components at -1.7, +14.3 and +32.0 km s$^{-1}$. The last one coincides with the stellar displacement, suggesting that the star may be associated with it, although the relatively large southern Galactic latitude (see Tab.~\ref{obstab}) render this possibility less likely if the interstellar lines are produced in the Galactic Plane.  Fig.~\ref{nai} displays in velocity space the positions of the \ion{Na}{i} lines and the nearby \ion{He}{i} 5876 \AA line. We see the perfect agreement between the velocity of the star and that of the last absorbing component. We conclude that the star is at the same distance than the absorbing cloud. From the Galactic rotation curve and using the parameters from \cite{reid14} we estimate the distance to IRAS 18131-3008 to be ~4.0 kpc. This estimation is close to the lower limit of the Gaia distance and agrees with the suggestion of d $<$ 7 kpc by \cite{savage17} , based on the agreement of the radial velocities of the ISM and the star.  
The map of \cite{chen19} gives the extinction in the Gaia bandpasses for the line of sight of IRAS 18131-3008, with E(B$_p$-R$_p$)= 0.48 for its maximum distance of 6.0 kpc, consistent with the value we obtain from the (B-V) in Tab~\ref{stardata} (0.50) and that in Gaia DR2 (0.58). 
The map is of litte help to distinguish between the short and large distances, because the extinction it is practically saturated from 4.0 kpc. However, a small increase in the extinction between 4.0 and 4.5 kpc and the smaller \cite{chen19} value favour the larger distance. We keep both values in our calculations, but we do not consider uncertainties in the short distance one, as it is just a rough estimation.

   In the case of IRAS 19336-0400 the \ion{Na}{i} lines only display the near absorption component at $\sim -6$ km s$^{-1}$, indicating that the line of sight leaves the Galactic disc very early in the direction of IRAS 19336-0400, which is a high velocity star (+67.4 \kms). 
The line of sight of this star is included in the extinction map by \cite{green18} (with a maximum angular resolution of 3.4 arcmin and a depth of 250 pc), but the extinction saturates for the corresponding pixel at a distance of 2.0 kpc with a value E(g$_{p1}$-r$_{p1}$)= 0.44\footnote{with g$_{p1}$ and r$_{p1}$ the magnitudes in the PanSTARRS-1 bandpasses, \cite[see e.g.][]{green18}}. This value is smaller than the one we get from the photometry in Tab.~\ref{stardata} (adopting the reddening vector in Table 1 of \cite{green18}). Therefore we keep the distance from \cite{bailer18}.
         
 For LSE 45 and LSE 163 stars the \ion{Na}{} spectrum is not so useful and the extinction maps do not include them. Contrary to IRAS 18131-3008, the \ion{Na}{i} interestellar lines in LSE 45 have a radial velocity pattern completely different from that of the stellar spectrum. The \ion{Na}{i} lines have components at $-3.3$, $-16.1$, $-37.6$ and $-55.8$ km s$^{-1}$, while the stellar spectrum shows a displacement of $+66$ km s$^{-1}$ with respect to the first \ion{Na}{i} line. This indicates that we are in the presence of a star away from the Galactic plane, consistently with its galactic latitude {\em l}= +11.4.  Finally, there are no interstellar lines in the spectrum of LSE 163 
Thus we have to rely on the \cite{bailer18} distances for these objects.

We also compare the optical and 2MASS photometry to obtain R$_{5495}$ (similar to R$_V$, see \cite{maiz14}), that we will use later to derive absolute visual magnitudes. We have used the extinction law from \cite{maiz14} to calculate the (B-V) corresponding to the 2MASS photometry and compare it with the observed (B-V) photometry given in Tab~\ref{stardata}. Table~\ref{bvjk} gives the differences between these two values for R$_{5495}$= 2.8. We get agreement between the optical and NIR photometry for all objects except LSE 163, that shows a marginal difference, which may be due to the impact of the fast rotation on the colors \cite[e.g.][]{townsend04}, but is too small to affect significantly to the discussion in the next section. Thus we adopt R$_{5495}$= 2.8 for all our objects (\cite{maiz18} obtain similar values for many of their objects). Again, the choice has not a significant effect on the discussion in next section. 

\begin{table*}
  \caption[]{Differences in the observed (B-V) minus the same value calculated from the 2MASS photometry using the \cite{maiz14} extinction law with R$_{5495}$= 2.8. Only in the case of LSE 163 the difference is slightly larger than the uncertainties.}
     \label{bvjk}
     \begin{tabular}{ccccc}
     \hline
     IRAS 17460-3114 & IRAS 18131-3008 & LSE 45 & IRAS 19336-0400 & LSE 163 \\
     0.063$\pm$0.094 & -0.047$\pm$0.094 & 0.113$\pm$0.117 & -0.155$\pm$0.192 & 0.111$\pm$0.084 \\
     \hline
     \end{tabular}
\end{table*}

  \begin{figure}
   \centering
   \includegraphics[bb=0 0 730 600, scale=0.35,angle=180]{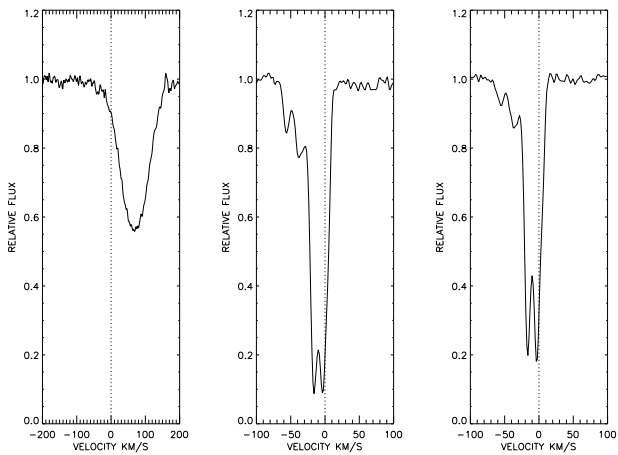}
      \caption{Same as Fig.~\ref{nai}, now for LSE 45. In this case, the radial velocities of the stellar and interstellar lines are completely different.}
         \label{nai2}
   \end{figure}

To derive absolute visual magnitudes, we obtain the extinction by calculating the intrinsic color from our best-fit model atmosphere and then comparing with the observed (B-V) color and making use of the apparent V-manitude (see Tab.~\ref{stardata}). We adopt R$_{5495}$= 2.8. The radius is obtained by integrating the spectral energy distribution of the best-fit model atmosphere in the optical range, so that the absolute visual magnitude is reproduced. Luminosities and masses (we call these spectroscopic masses) are then derived from the standard formulae. Finally, we place the stars on the Hertzsprung-Russell Diagram and read evolutionary mass, i.e., the mass predicted by the evolutionary tracks at that position \citep[we use][for massive and post-AGB stars respectively]{ekstroem12, bertolami16}. The values are given in Tab.~\ref{dr2datab}. Errors include the uncertainty in the distances from \cite{bailer18} and in the stellar parameters (Tab.~\ref{stparam}).

 \begin{table*}
 \caption[]{DR2 data for the stars in this work. Parallaxes from \cite{gaia18} are in miliarcsec and distances from \cite{bailer18} in pc. The Renormalized Unit Weight Error (RUWE) is defined in \cite{lindegren18}, who recommend a value below 1.40 for accurate measurements. z gives the height above or below the Galactic plane.}
\label{dr2data}
\begin{tabular}{lcccc}
\hline
Star        &      $\pi$ (mas)                  & RUWE   &      d (pc)      &  z (pc)  \\
\hline
IRAS 17460-3114 &  0.782$\pm$0.077 & 0.79 & 1249$^{+141}_{-114}$       &  20$\pm$2   \\ 
IRAS 18131-3008  & -0.008$\pm$0.073 &  0.60 & 8934$^{+4168}_{-2685}$  & 498$^{+232}_{-150}$  \\
                               &                               &          &  4000                                 & 223                         \\
LSE 45                   &  0.219$\pm$0.084 &  0.98 & 3713$^{+1471}_{-1890}$  & 370$^{+147}_{-188}$  \\
IRAS 19336-0400 &  0.163$\pm$0.053 &  1.61 & 4746$^{+1450}_{-956}$    & 475$^{+145}_{-96}$   \\
LSE 163                 &  0.235$\pm$0.076  &  1.17 & 3417$^{+1062}_{-698}$    & 592$^{+184}_{-121}$  \\ 
\hline
\end{tabular}
\end{table*}

  \begin{table*}
 \caption[]{Stellar parameters derived in this work. E(B-V) is derived from the photometry in Tab.~\ref{stardata} and the intrinsic (B-V)$_0$ values from the final model atmospheres. Absolute magnitudes, radii, luminosities and masses are derived as explained in text. M$_{sp}$ are spectroscopic masses and M$_{ev}$ evolutionary masses.}
\label{dr2datab}
\begin{tabular}{lccccccc}
\hline
Star        &   E(B-V) & A$_V$ & M$_{\rm V}$ &  R/R$_\odot$  &  log L/L$_\odot$  & M$_{sp}$/M$_\odot$ & M$_{ev}$/M$_\odot$ \\
\hline
IRAS 17460-3114 &   0.527$\pm$0.054 & 1.476$\pm$0.151& -4.11$^{+0.25}_{-0.29}$ & 7.0$^{+0.9}_{-0.8}$    & 4.92$^{+0.12}_{-0.10}$  & 25.1$^{+9.8}_{-9.2}$ & 24.6$\pm$1.2 \\ 
IRAS 18131-3008 &  0.158$\pm$0.028 & 0.442$\pm$0.078 & -6.17$^{+0.66}_{-1.02}$ & 21.0$^{+9.9}_{-6.4}$ & 5.50$^{+0.40}_{-0.26}$ & 32.3$^{+30.5}_{-20.8}$ & 32.5$^{15.2}_{-8.3}$ \\
                              &         &                    & -4.47                                & 10.4                            & 4.90                               &   7.89                             &                         \\
LSE 45                 &  0.249$\pm$0.097 & 0.697$\pm$0.272 & -2.73$^{+1.14}_{-0.91}$ & 4.5$^{+1.9}_{-2.3}$     & 4.11$^{+0.36}_{-0.46}$ & 2.4$^{+2.1}_{-2.7}$ & 0.69$^{+0.21}_{-0.23}$ \\
IRAS 19336-0400 &  0.632$\pm$0.168 & 1.769$\pm$0.470 & -2.05$^{+0.65}_{-0.82}$ & 3.3$^{+1.3}_{-1.0}$     & 3.79$^{+0.33}_{-0.27}$ & 1.2$^{+1.4}_{-0.9}$ & 0.58$^{+0.12}_{-0.05}$ \\
LSE 163                 &  0.118$\pm$0.047 & 0.330$\pm$0.132 & -2.54$^{+0.46}_{-0.69}$ & 4.9$^{+1.6}_{-1.0}$     & 3.81$^{+0.28}_{-0.18}$ & 2.9$^{+2.0}_{-1.4}$ & 0.58$^{+0.11}_{-0.05}$ \\ 
\hline
\end{tabular}
\end{table*}

  \section{Discussion}
  
  We discuss here the nature of our sample of hot stars, particularly whether they are massive young stars or CSPN in a low excitation stage, which implies that the central object is still contracting and becoming hotter. The abundances we obtained in the previous section do not depart from solar abundances (except the Si abundance of IRAS 19336-0400 and mildly in the case of CNO in LSE 163) and cannot be used to distinguish  between both possibilities. However, the combination of Gaia DR2 data with the stellar parameters derived in this work will allow us to make this distinction. Although uncertainties are large, mainly due to the large distances of most of the targets, they are sufficient to separate massive supergiants from the much smaller CSPN. 
  
\begin{figure}
   \centering
     \includegraphics[bb=0 0 700 650,scale=0.38,angle=180]{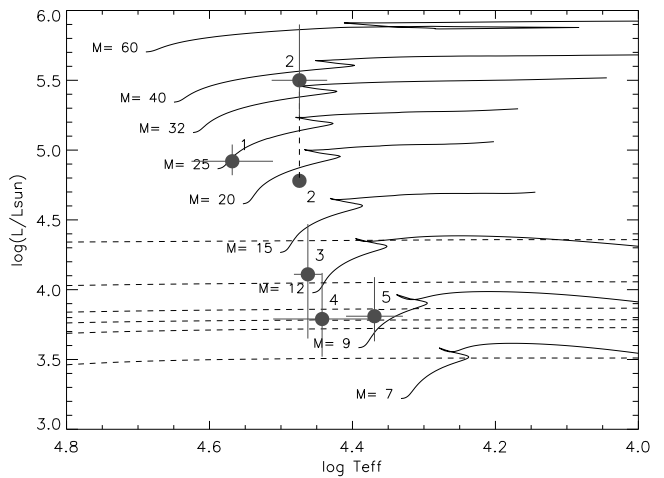}
      \caption{Hertzsprung-Russell Diagram showing the evolutionary tracks for stars with initial masses between 7 and 60 M$_\odot$ \citep[][solid lines]{ekstroem12} and from for post-AGB stars with initial (final) masses (decreasing from top to bottom) of 1.0 (0.528), 1.25 (0.561), 1.5 (0.576), 2.0 (0.580), 3.0 (0.657) and 4.0 (0.833) M$_\odot$ \citep[][dashed lines]{bertolami16}, which are nearly horizontal lines in the plot scale. The masses given in the plot are initial masses for the massive stars tracks. Our targets are shown as solid dots with errors, and the numbers close to them label the stars: (1): IRAS 17460-3114; (2): IRAS 18131-3008; (3): LSE 45; (4): IRAS 19336-0400; (5): LSE 163. Two values are shown for IRAS 18131-3008, corresponding to the two distances in Tab.~\ref{dr2data}}
         \label{hrd}
   \end{figure}

  Figure~\ref{hrd} shows the position of our stars in the Hertzsprung-Russell Diagram (HRD), together with the evolutionary tracks from \cite{ekstroem12} for massive stars with initial masses between 9 and 60 M$_\odot$ (we have added the track for 7 M$_\odot$ for completeness) and \cite{bertolami16} for post-AGB stars with initial masses between 1.0 and 4.0 M$_\odot$. IRAS 17460-3114 and IRAS 18131-3008 fall in the zone of massive stars, whereas the other three (LSE 45, IRAS 19336-0400 and LSE 163) fall in  regions that could correspond to either less massive stars at the beginning of their evolution or to stars in the early phases of post-AGB evolution.
  
  IRAS 17460-3114 (HD 161853) is a binary star recently studied in detail by \cite{gamen15}. They derive a distance of 1.35$\pm$0.25 kpc that agrees very well with the distance of 1.25$\pm$0.14 kpc from DR2 (see Tab.~\ref{dr2data}). Correcting from interstellar extinction using the final model atmosphere to calculate the intrinsic (B-V) color, this indicates an absolute magnitude of M$_v$=  -4.11$^{+0.25}_{-0.29}$ and a luminosity of log(L/L$_\odot$)= 4.92$^{+0.11}_{-0.09}$, in agreement with the values given by \cite{martins05} for an O8V star. The spectroscopic and evolutionary masses quoted in Tab.~\ref{dr2datab} agree with each other and indicate a massive star. The rotational velocity is also larger than typical for hot post-AGB stars \citep[see e.g.][]{mello12}. The Galactic latitude of IRAS 17460-3114 is -0.19 degrees (see Tab.~\ref{obstab}) and the star is associated to an H\,{\sc ii} region \citep[see][and references therein]{gamen15}. Although the object was included as a hot post-AGB star by \cite{mello12} in their sample based on the IRAS colours, they leave their nature undefined and it also appears in the Torun catalog of disqualified post-AGB stars by \cite{szczerba07}.  In agreement with the latter authors and \cite{gamen15} we conclude that this is a young massive star in the Galactic plane, which also agrees with the lack of nebular lines that should be already present in a CSPN at this high temperature.
    
  IRAS 18131-3008 (HD 167402) has a very low parallax, indicating a large distance (which was not the case in DR1 data). Adopting the distance from \cite{bailer18} quoted in Tab.~\ref{dr2data} results (after correcting from extinction in the same way as for IRAS 17460-3114) in an absolute magnitude of M$_v$= -6.17$^{+0.66}_{-1.02}$ and a luminosity of  log(L/L$_\odot$)= 5.50$^{+0.40}_{-0.26}$, corresponding to a large radius (R/R$_\odot$= 21.0$^{+9.9}_{-6.4}$, see Tab-~\ref{dr2datab}). The spectroscopic mass quoted in the table is consistent with the predictions from evolutionary tracks. These values correspond to a massive and luminous OB supergiant, although the large distance uncertainty makes this argument weaker, as the distance might be smaller as we have indicated in the previous section. Adopting the distance of 4.0 kpc, that we have considered a lower limit, we obtain parameters corresponding to a less massive evolved star. In this case the mass we obtain is much lower than the evolutionary one, favouring the far distance. Moreover, to reduce the absolute magnitude to values typical for CSPN (M$_v \gtrapprox$ -3) we would need a distance comparable to that of IRAS 17460-3114 ($\lessapprox$ 2 kpc) which should give a measurable parallax. Therefore, in spite of its relatively large southern Galactic latitude (-6.4 degrees) and large distance from the Galactic plane we conclude that this object is a massive B-supergiant or bright giant.
        
   LSE 45 (CD-49 8217) has a parallax indicating a distance of $\approx$3.7 kpc (see Tab.~\ref{dr2data}), which results in an absolute magnitude of -2.73$^{+1.14}_{-0.91}$ and a luminosity of log(L/L$_\odot$)= 4.11$^{+0.36}_{-0.46}$, again correcting for extinction using the synthetic photometry calculated with our best-fit model atmosphere. The radius is R= 4.5$^{+1.9}_{-2.3}$  R$_\odot$ and the spectroscopic mass is 2.4$^{+2.1}_{-2.7}$ M$_\odot$, too large for a post-AGB star. The position in the HRD is consistent with a post-AGB star, with an evolutionary mass of 0.69$^{+0.21}_{-0.23}$ M$_\odot$. Although the error bars overlap, this introduces some tension in the stellar mass. A similar difficulty is found when we use the Kiel diagram or the spectroscopic HRD \citep{langer14}, that are independent of distance. This is reminiscent of the mass discrepancy, a well known problem in massive stars, not completely solved \citep[see][and references therein]{h92, mokiem07, markova18}.The reason for this inconsistency (f.e., possible variability, contaminated magnitude, wrong distance, extra line broadening...) is not clear from our study and will require new observations or analyses. The data however are less compatible with a slightly evolved massive star of this temperature and luminosity. The possibility that LSE 45 was originally a low mass star that has accreted material from a massive companion and has later been ejected from a binary system is consistent with the large radial velocity (Tab.~\ref{stardata}), but is not supported by the normal CNO abundances or the low rotational velocity \citep[see][although these authors study systems with higher secondary masses, the effects would be similar for lower mass stars]{renzo19}. Data might thus be considered to favour the possibility of a young CSPN still contracting and with a relatively low effective temperature that cannot yet excite the material around the star, as was suggested by \cite{gauba03}. But then the reason why we see the nebular lines in IRAS 19336-0400 (with a similar temperature, see next paragraph) and the too large spectroscopic mass will require further examination.
 
    IRAS 19336-0400 (SS441) is at low Galactic latitude (-11.7 degrees). Its low parallax indicates a distance of $\approx$4.7 kpc, and a height of nearly 0.5 kpc below the Galactic plane. This translates into an absolute magnitude of -2.05$^{+0.65}_{-0.82}$  and a luminosity of log(L/L$_\odot$)= 3.79$^{+0.33}_{-0.27}$. The radius we obtain is R= 3.3$^{+1.3}_{-1.0}$ R$_\odot$ and the spectroscopic mass of only 1.2$^{+1.4}_{-0.9}$ M$_\odot$ agrees within uncertainties with the evolutionary one, of 0.58$^{+0.12}_{-0.05}$ M$_\odot$. Although the Oxygen abundance is close to solar, Silicon, with a smaller uncertainty, indicates a slightly metal-poor star. These data are again consistent with a young CSPN still contracting (as gravity is low) and agree with the observation of nebular lines in the Balmer profiles (much weaker in the \ion{He}{i} lines) and those by \cite{gledhill15}, who observed H and He recombination lines in the NIR spectrum, indicating the advanced formation of a Planetary Nebula. Its effective temperature is similar to that of LSE 45, that does not show nebular lines. We do not know the reason for this difference, although we speculate that it could be due to small differences in the stellar parameters (in which case LSE 45 should show nebular lines in a short time in terms of evolution) or to differences in the gas properties, or both. 
           
   LSE 163 (CD-42 8141): the quoted uncertainties in the parallaxes (see Tab.~\ref{dr2data}) allow to discriminate between a massive supergiant and a lower mass star. The star is at high Galactic latitude (+19.3 degrees) and its parallax places it at a distance of $\approx$3.4 kpc, implying a distance of about 0.6 kpc to the Galactic plane, an absolute magnitude of -2.54$^{+0.46}_{-0.69}$ and a luminosity of log(L/L$_\odot$)= 3.81$^{+0.28}_{-0.18}$. Correcting the derived gravity for centrifugal acceleration we obtain a mass of 2.9$^{+2.0}_{-1.4}$ M$_\odot$, which is not consistent with the mass predicted by evolutionary tracks for post-AGB stars at this effective temperature and luminosity (similarly to LSE 45, see above, but now with an even stronger effect due to the larger spectroscopic mass and smaller uncertainties). Moreover, the rotational velocity (259$\pm$15 \kms) is very large compared to typical velocities of hot post-AGB stars \citep[clearly less than 100 \kms, see][]{mello12, prinja12}. Although some authors like \cite{geier13} or \cite{DeMarco15} have suggested that fast rotating sdB or CSPN stars may be the result of mergers, the rotational velocities of the objects they analyzed are much lower than that of LSE 163. The combination of a very high rotational velocity, high Galactic latitude, slightly large radial velocity and mild CNO enhancement is suggestive of a history of binary interaction. LSE 163 then would be the result of interaction in a binary system where it gained mass and angular momentum from a massive companion that exploded \citep[see][and comments in the LSE 45 paragraph]{renzo19}, although in this scenario it is difficult to explain the high temperature at the surface. Alternatively, we may consider LSE 163 being a mildly stripped star, following the recent work by \cite{gotberg18}. These authors have presented models of stripped stars produced by the interaction in a binary systems. The products resulting from such interaction present a continous of spectral characteristics from sdB up to Wolf-Rayet. However these models predict a much larger gravity for a solar metallicity stripped star of this effective temperature and the CNO abundances should reflect a higher contamination.
     
 \section{Conclusions}

We have analyzed high-resolution spectra of five hot objects to clarify their possible nature as post-AGB stars. We have derived their radial, rotational and macroturbulent velocities. The most relevant results is the large rotational velocity of LSE 163, much larger than expected from its spectral classification. We have also derived their stellar parameters and CNO and Si abundances using spherical NLTE model atmospheres with mass-loss and automatic methods. Stellar parameters are consistent with the spectral classifications. The abundances are solar within uncertainties in all cases, with the exception of a small Si abundance (but not O) in IRAS 19336-0400 and a mild CNO contamination in LSE 163. Thus abundances did not help much to clarify the possible nature as post-AGB objects.

We have used the distances from \cite{bailer18} (who used Gaia DR2 parallaxes) to derive absolute magnitudes. Most parallaxes are small indicating large distances. The DR2 data for our targets have a Renormalized Unit Weight Error (RUWE) smaller than 1.40 (which indicates reliable data) with the exception of IRAS 19336-0400, for which it is 1.61, although their relative errors in parallax and distance are only 30$\%$. We have used the absolute magnitudes to derive radii and luminosities. Absolute magnitudes, radii and luminosities are very useful in separating massive and low-mass stars. We find that IRAS 17460-3114 and IRAS 18131-3008 (the latter in spite of its height below the Galactic plane) have values that are consistent with those expected for massive OB stars of their spectral type and luminosity class. In the first case, this confirms the results of other researchers \citep[see][and references therein]{gamen15}. On the other hand, LSE 45, IRAS 19336-0400 and LSE 163 have very small radii and masses for massive supergiants. Our results confirm that IRAS 19336-0400 is a hot post-AGB star, in agreement with the work of \cite{gledhill15} who detect near infrared recombination lines in IRAS 19336-0400. However, we cannot confirm the nature of LSE 45 and LSE 163 as hot post-AGB stars still contracting but without showing yet nebular lines, as suggested by \cite{gauba03}. In both cases, the derived spectroscopic masses  are too large compared with those predicted by the post-AGB evolutionary tracks of \cite{bertolami16} but much more incosistent with those of a massive early B star of any luminosity class. In the case of LSE 163, the combination of very high rotational velocity, high Galactic latitude, slightly large radial velocity and mild CNO enhancement suggests a possible history of binary interaction, although not all observed parameters can be explained simultaneously by the different scenarios.
  
\section{acknowledgements}
We would like to acknowledge Dr. R. M\'endez, Dr. O. De Marco, Dr. M. Garcia and Dr. D.J. Lennon for helping us with different aspects of the paper. The two referees of the paper are also acknowledged, as their comments contributed to improve it.
This paper is based on data products from observations made with ESO Telescopes at the La Silla Paranal Observatory under programme ID 077.D-0478A.
MP is thankful to Prof. Wako Aoki and  Director General of National Astronomical Observatories of Japan (NAOJ), Prof. Saku Tsuneta, for their kind encouragement and support. 
AH and SSD acknowledge that part of this work is funded by the Spanish MINECO under grants PGC-2018-091\,3741-B-C22 and SEV 2015-0548, and the Agencia Canaria de Investigaci—n, Innovaci—n y Sociedad de la Informaci—n del Gobierno de Canarias (ACIISI) and the European Regional Development Fund (ERDF) under grant with reference ProID2017010115. This work has made use of data from the European Space Agency mission Gaia (https://www.cosmos.esa.int/gaia), processed by the Gaia Data Processing and Analysis Consortium (DPAC, https://www.cosmos.esa.int/web/gaia/dpac/consortium). Funding for the DPAC has been provided by national institutions, in particular the institutions participating in the Gaia Multilateral Agreement.    
This work used the CONDOR workload management system (http://www.cs.wisc.edu/condor/) implemented at the Instituto de Astrof\'isica de Canarias for the calculation of the model atmosphere grids.

\appendix
\section{Line Equivalent Widths}

 \begin{table*}
 \caption[]{Carbon measured equivalent widths in m\AA. \ion{C}{iv} lines and those marked with an asteric were not used for the abundance analysis. All line wavelengths are in \AA}
\label{tabewsc}
\begin{tabular}{lrrrrr}
\hline
Line  & IRAS           & IRAS           & LSE 45 & IRAS       & LSE 163 \\
        & 17460-3114 &   18131-3008 &          & 19336-0400 &           \\
\hline
C {\sc II} 3918 & ---                  &  ---                & ---                & ---                  & 150$\pm$44 \\
C {\sc II} 3920 & 109$\pm$29  &  19$\pm$11  & ---                & ---                  & 121$\pm$7 \\
C {\sc II} 4267 & ---                  &  158$\pm$12& 159$\pm$8  & ---                  & 314$\pm$57 \\
C {\sc II} 5133 & ---                  &  ---                &  33$\pm$14  & ---                  & ---                \\
C {\sc II} 5145 & ---                  &  ---                &  39$\pm$9    & ---                  & ---                \\
C {\sc II} 5151 & ---                  &  ---                &  24$\pm$6    & ---                  & ---                \\
C {\sc II} 6098 & ---                  &  ---                &  26$\pm$13*  & ---                  & ---                \\
C {\sc II} 6578 & ---                  &  ---                & ---                & ---                   & 376$\pm$66* \\
C {\sc II} 6582 & ---                  &  ---                &  61$\pm$21  & ---                  & 188$\pm$34 \\
\hline
C {\sc III} 4325 & ---                 & 57$\pm$17   & ---                 & ---                  & ---                \\
C {\sc III} 4647 & 170$\pm$13 & ---                 & 192$\pm$52 & 388$\pm$47 & ---                \\
C {\sc III} 4650 &  84$\pm$5    & 326$\pm$51 & ---                 & 288$\pm$29 & ---                \\
C {\sc III} 4651 &  99$\pm$4    & 244$\pm$19 & ---                 & 240$\pm$19 & ---                \\
C {\sc III} 4665 & ---                 & 72$\pm$12   & 46$\pm$13   & ---                  & ---                \\
\hline
C {\sc IV} 5801 & 363$\pm$44 & 119$\pm$16 & ---                 & ---                  & ---                \\
C {\sc IV} 5811 & 220$\pm$46 & 74$\pm$12   & ---                 & ---                  & ---                \\
\hline
\end{tabular}
\end{table*}

 \begin{table*}
 \caption[]{Nitrogen measured equivalent widths in m\AA. Lines marked with an asteric were not used for the abundance analysis. For IRAS 17460-3114 we measured the combined equivalent width of the N{\sc III} 4511-15 \AA lines. All line wavelengths are in \AA}
\label{tabewsn}
\begin{tabular}{lrrrrr}
\hline
Line     & IRAS           & IRAS           & LSE 45 & IRAS       & LSE 163 \\
            & 17460-3114 &   18131-3008 &          & 19336-0400 &           \\
\hline
N {\sc II} 3995 & ---                 &  96$\pm$13*   &  99$\pm$12  &  ---                & ---               \\
N {\sc II} 4447 & ---                 &100$\pm$29   &  ---                &  ---                & ---               \\
N {\sc II} 4630 & ---                 &  85$\pm$14   &  69$\pm$5    &  ---                & ---               \\
N {\sc II} 4779 & ---                 &  23$\pm$8     &  ---                &  ---                & ---               \\
N {\sc II} 4803 & ---                 &  41$\pm$13   &  59$\pm$10*  &  ---                & ---               \\
N {\sc II} 4994 & ---                 & ---                  &  25$\pm$11*  &  ---                & ---               \\
N {\sc II} 5007 & ---                 &  ---                 &  ---                &  ---                &  73$\pm$11 \\
N {\sc II} 5001 & ---                 &  96$\pm$13   &  82$\pm$3*   &  ---                & 102$\pm$24 \\
N {\sc II} 5002 & ---                 &  ---                 &  ---                &  ---                & 130$\pm$16 \\
N {\sc II} 5005 & ---                 &  48$\pm$12*   &  56$\pm$11 &  ---                & 147$\pm$18 \\
N {\sc II} 5045 & ---                 &  48$\pm$8*     &  37$\pm$9*   &  ---                & 114$\pm$17* \\
N {\sc II} 5495 &  66$\pm$42  & ---                  &  ---                &  ---                & ---               \\
N {\sc II} 5666 & ---                 & ---                  &  37$\pm$11  &  ---                & ---               \\
N {\sc II} 5676 & ---                 & ---                  &  21$\pm$4    &  ---                & ---               \\
N {\sc II} 5679 & ---                 &  69$\pm$10*   &  80$\pm$18  &  ---                & ---               \\
N {\sc II} 5710 & ---                 &  32$\pm$11   &  ---                &  ---                & ---               \\
N {\sc II} 6482 & ---                 & ---                  &  53$\pm$18*  &  ---                & ---               \\
\hline
N {\sc III} 3998 & 18$\pm$9    &  ---                 &  ---                &  ---                & ---               \\
N {\sc III} 4003 & 14$\pm$8    &  ---                 &  ---                &  ---                & ---               \\
N {\sc III} 4097 & 211$\pm$12 & 153$\pm$12 &  ---                &  ---                & 189$\pm$77 \\
N {\sc III} 4195 & ---                 &  34$\pm$7    &  ---                &  ---                & ---               \\
N {\sc III} 4379 & 89$\pm$18*  &  ---                 &  ---                &  ---                & ---               \\
N {\sc III} 4510 & ---                 &  75$\pm$19  &  ---                &  ---                & ---               \\
N {\sc III} 4511 & ---                 &  ---                 &  53$\pm$13 &  ---                &  ---               \\
N {\sc III} 4515 & ---                 & 137$\pm$31 &  36$\pm$12 &  ---                & ---               \\
N{\sc III} 4511-15 & 109$\pm$22 &                  &                     &                      &                  \\  
N {\sc III} 4523 & ---                 &  24$\pm$15*  &  ---                &  ---                & ---               \\
N {\sc III} 4535 & ---                 &  ---                 &  56$\pm$17* &  ---                &  ---               \\
N {\sc III} 4634 & ---                 & 120$\pm$18* &  53$\pm$11  &  ---                & ---               \\
N {\sc III} 4640 & 28$\pm$11   &  75$\pm$19  &  ---                &  ---                & ---               \\
N {\sc III} 4641 & ---                 & 125$\pm$20 & 141$\pm$25 &  67$\pm$20* & ---               \\
N {\sc III} 4867 & 386$\pm$16 & ---                 & ---                 & ---                 & 377$\pm$122 \\
\hline
\end{tabular}
\end{table*}
   
 \begin{table*}
 \caption[]{Oxygen measured equivalent widths in m\AA. \ion{O}{iii} lines and lines marked with an asteric were not used for the abundance analysis. All line wavelengths are in \AA}
\label{tabewso}
\begin{tabular}{lrrrrr}
\hline
Line     & IRAS           & IRAS           & LSE 45 & IRAS       & LSE 163 \\
            & 17460-3114 &   18131-3008 &          & 19336-0400 &           \\
\hline
O {\sc II} 3911 & ---                 &  95$\pm$12*    & 113$\pm$11 & ---                   & ---                   \\
O {\sc II} 3945 & ---                 &  66$\pm$14   &  77$\pm$16 & ---                   & ---                   \\
O {\sc II} 3954 & ---                 &107$\pm$21   & ---                & ---                   & ---                   \\
O {\sc II} 3982 & ---                 &  36$\pm$5     &  60$\pm$11 & ---                   & ---                   \\
O {\sc II} 4069 & 301$\pm$52 & 457$\pm$55* & 225$\pm$46 & ---                   & 124$\pm$26* \\
O {\sc II} 4072 & ---                 & ---                  & 133$\pm$20 & ---                   & 150$\pm$32* \\
O {\sc II} 4076 &  40$\pm$16 & 172$\pm$38   & 190$\pm$34 & ---                   & 166$\pm$35 \\
O {\sc II} 4078 & ---                 & ---                  & 59$\pm$16  & ---                   & ---                   \\
O {\sc II} 4086 & ---                 &  82$\pm$12*   & 146$\pm$42 & ---                   & ---                   \\
O {\sc II} 4089 & 167$\pm$2  & 103$\pm$7*     & 81$\pm$5*  & 132$\pm$29* & ---                   \\
O {\sc II} 4129 & ---                 & ---                   & 38$\pm$7  & ---                   & ---                   \\
O {\sc II} 4132 & ---                 &  92$\pm$19    & 95$\pm$18  & ---                   & ---                   \\
O {\sc II} 4153 & ---                 & 130$\pm$9*     & 168$\pm$13 & ---                   & ---                   \\
O {\sc II} 4156 & ---                 &  20$\pm$3*      & ---                & ---                   & ---                   \\
O {\sc II} 4185 & ---                 &  236$\pm$10*    &  88$\pm$41 & ---                   & ---                   \\
O {\sc II} 4189 & ---                 &  82$\pm$12    & 107$\pm$10 & ---                  & ---                   \\
O {\sc II} 4317 & ---                 & 109$\pm$21  & 121$\pm$25 & ---                   & ---                   \\
O {\sc II} 4319 & ---                 & 106$\pm$13  & 135$\pm$16 & ---                   & ---                   \\
O {\sc II} 4351 & ---                 &  139$\pm$10    & 226$\pm$ 10* & ---                   & ---                   \\
O {\sc II} 4366 & 181$\pm$20 & 139$\pm$21  & 128$\pm$15 & ---                   & ---                   \\
O {\sc II} 4414 &  52$\pm$19 & 160$\pm$11  & 177$\pm$26* & ---                   & 177$\pm$62 \\
O {\sc II} 4416 & ---                 & 124$\pm$17  & 146$\pm$36 & ---                   & 150$\pm$53 \\
O {\sc II} 4452 & ---                 &  40$\pm$16   &  64$\pm$9  & ---                   & ---                   \\
O {\sc II} 4638 & ---                 & 150$\pm$18  & 125$\pm$30 & ---                   & ---                   \\
O {\sc II} 4641 & ---                 & ---                  & 98$\pm$5   &       79$\pm$14 & ---                   \\
O {\sc II} 4650 &  65$\pm$3   & 263$\pm$18  & ---                &     383$\pm$8 & ---                   \\
O {\sc II} 4661 & ---                 & 128$\pm$16  & 162$\pm$14 & ---                   & ---                   \\
O {\sc II} 4673 & ---                 &  16$\pm$39*   &  36$\pm$5  & ---                   & ---                   \\
O {\sc II} 4676 & ---                 & 141$\pm$23  & 129$\pm$16 & ---                   & ---                   \\
O {\sc II} 4696 & ---                 &  ---                 & 21$\pm$9  & ---                   & ---                   \\
O {\sc II} 4699 & ---                 &  80$\pm$0      & ---                & ---                   & ---                   \\
O {\sc II} 4891 & ---                 &  28$\pm$11    &  41$\pm$11 & ---                   & ---                   \\
O {\sc II} 4906 & ---                 &  63$\pm$9      &  64$\pm$9 & ---                   & ---                   \\
O {\sc II} 4941 & ---                 &  45$\pm$5      &  50$\pm$16 & ---                   & ---                   \\
O {\sc II} 4943 & ---                 &  75$\pm$7      &  86$\pm$16 & ---                   & ---                   \\
O {\sc II} 4956 & ---                 &  17$\pm$13    & ---                & ---                   & ---                   \\
O {\sc II} 5160 & ---                 &  53$\pm$12*    &  51$\pm$11 & ---                   & ---                   \\
O {\sc II} 5207 & ---                 &  48$\pm$15    &  58$\pm$11 & ---                   & ---                   \\
O {\sc II} 6721 & ---                 &  47$\pm$27   & 80$\pm$17 & ---                   & ---                   \\
\hline
O {\sc III} 3759 & 180$\pm$41 & ---                  & 111$\pm$27 & ---                  & ---                   \\
O {\sc III} 3961 & ---                 &  91$\pm$16  &  30$\pm$14  & ---                   & ---                   \\
O {\sc II} 5508 & ---                  &  30$\pm$14  & ---                 & ---                   & ---                   \\
O {\sc III} 5592 & 195$\pm$22 & 173$\pm$17 &  59$\pm$14  & 142$\pm$59 & ---                   \\
\hline
\end{tabular}
\end{table*}

 \begin{table*}
 \caption[]{Silicon measured equivalent widths in m\AA. Lines marked with an asteric were not used for the abundance analysis. All line wavelengths are in \AA}
\label{tabewssi}
\begin{tabular}{lrrrrr}
\hline
Line     & IRAS           & IRAS           & LSE 45 & IRAS       & LSE 163 \\
           & 17460-3114 &   18131-3008 &          & 19336-0400 &           \\
\hline
Si {\sc III} 3791 & 307$\pm$33* & 104$\pm$29* &  65$\pm$20* & ---                  & ---                 \\
Si {\sc III} 3806 & ---                 & ---                  & 141$\pm$6  & ---                  & 464$\pm$112 \\
Si {\sc III} 4552 & ---                 & 247$\pm$17 & 283$\pm$12 & 285$\pm$54 & 475$\pm$90  \\
Si {\sc III} 4567 & ---                 & 239$\pm$18 & 240$\pm$28 & 222$\pm$82 & 305$\pm$62 \\
Si {\sc III} 4574 & ---                 & 129$\pm$16 & 136$\pm$13 & ---                  & ---                 \\
Si {\sc III} 4716 & ---                 & ---                 & 19$\pm$4     & ---                  & ---                 \\
Si {\sc III} 4813 & ---                 &  38$\pm$8    & 39$\pm$10   & ---                  & ---                 \\
Si {\sc III} 4819 & ---                 &  56$\pm$8    & 67$\pm$9     & ---                  & ---                 \\
Si {\sc III} 4829 & ---                 &  48$\pm$6    & 58$\pm$9     & ---                  & ---                 \\
Si {\sc III} 5739 & ---                 & 147$\pm$45 & 176$\pm$11 & ---                  & 341$\pm$98 \\
\hline
Si {\sc IV} 4089 &  83$\pm$0   & 502$\pm$28* & 330$\pm$29 & 371$\pm$57* & ---                 \\
Si {\sc IV} 4116 &  72$\pm$16 & 347$\pm$95 & 257$\pm$15 & 325$\pm$76 & ---                 \\
Si {\sc IV} 4212 & ---                &  67$\pm$11  &  30$\pm$4    & ---                  & ---                 \\
Si {\sc IV} 4631 &  75$\pm$22 & 101$\pm$11 &  71$\pm$12 & ---                  & ---                 \\
Si {\sc IV} 4654 &  74$\pm$8  & 120$\pm$19 &  88$\pm$21  & ---                  & ---                 \\
Si {\sc IV} 6701 & ---                &  50$\pm$16* & ---                  & ---                  & ---                 \\
\hline
\end{tabular}
\end{table*}


\begin{thebibliography}{}

 \bibitem[Arkhipova et al. (2012)]{arkhipova12} Arkhiopova, V.P., Burlak, M.A., Esipov, V.F., Ikonnikova, N.P. \& Komissarova, G.V., 2012, AstL 38, 157

 \bibitem[Asplund et al.(2009)]{asplund09} Asplund, M., Grevesse, N., Sauval, A.~J., \& Scott, P.\ 2009, \araa, 47, 481
 
 \bibitem[Bailer-Jones (2015)] {bailer15} Bailer-Jones, C.A.L., 2015, PASP, 127, 994 

 \bibitem[Bailer-Jones et al. (2018)] {bailer18} Bailer-Jones, C.A.L., Rybizki, J., Fouesneau, M., Mantelet, G. \& Andrae, R., 2018, \aj, 156, A58 
 
 \bibitem[Chen et al. (2019)]{chen19} Chen, B.-Q., Huang, Y., Yuan, H.B. et al., 2019, MNRAS, 483, 4277
 
 \bibitem[Conlon et al. (1993)]{conlon93}Conlon E.~S., Dufton P.~L., McCausland R.~J.~H., Keenan F.~P., 1993, \apj, 408, 593 
  
 \bibitem[Conlon et al. (1991)]{conlon91} Conlon E.~S., Dufton P.~L., Keenan F.~P., McCausland R.~J.~H., 1991, \mnras, 248, 820 
 
 \bibitem[De Marco et al. (2015)]{DeMarco15} De Marco, O., Long, J., Jacoby, G.H. et al., 2015, \mnras 448, 3587
 
 \bibitem[Drilling \& Bergeron (1995)]{drilling95} Drilling, J.S., and Bergeron, L.E., 1995, PASP, 107, 846
 
 \bibitem[Dufton et al.(2011)]{dufton11} Dufton, P.L., Dunstall, P.R., Evans, C.J. et al., 2011, \apj, 743, L22

 \bibitem[Ekstroem et al.(2012)]{ekstroem12} Ekstroem, S., Georgy, C., Eggenberger, P. et al., 2012, A\&A, 537, A146

 \bibitem[Kaufer et al. (1999)]{kaufer99} Kaufer A., Stahl O., Tubbesing S., N{\o}rregaard P., Avila G., Francois P., Pasquini L., Pizzella A., 1999, Msngr, 95, 8 
 
\bibitem[Gaia Collaboration et al.(2018)]{gaia18} Gaia Collaboration, Brown, A.~G.~A., Vallenari, A., et al.\ 2018, arXiv:1804.09365

\bibitem[Gamen et al. (2015)]{gamen15} Gamen,R., et al., 2015, A\&A 584, A7
 
 \bibitem[Gauba et al. (2003)]{gauba03} Gauba G., Parthasarathy M., Kumar B., Yadav R.~K.~S., Sagar R., 2003, A\&A, 404, 305 
 
 \bibitem[Garrison et al. (1977)]{garrison77} Garrison R.~F., Hiltner W.~A., Schild R.~E., 1977, ApJS, 35, 111 
 
 \bibitem[Geier et al. (2013)]{geier13} Geier, S., Heber, U., Heuser, C. et al., 2013, A\&A 551, L4
 
 \bibitem[Gledhill \& Forde (2015)]{gledhill15} Gledhill, T.M. \& Forde, K.P. \ 2015, \mnras 447, 1080
 
 \bibitem[G\"otberg et al. (2018)]{gotberg18} G\"otberg, Y., de Mink, S.E., Groh, J.H., 2018, A\&A, 615, A78
 
 \bibitem[Green et al. (2018)]{green18} Green, G.M., Schlafly, E.F., Finkbeiner, D. et al., 2018, MNRAS, 478, 651

 \bibitem[Herrero et al.(1992)]{h92} Herrero A., Kudritzki R.~P., Vilchez J.~M., Kunze D., Butler K., Haser S., 1992, A\&A, 261, 209 
 
 \bibitem[Herrero et al. (2002)]{h02} Herrero, A., Puls, J. \& Najarro, F., 2002, \aap, 396, 949
 
 \bibitem[Hillier \& Miller(1998)]{HM98} Hillier, D.~J., \& Miller, D.~L.\ 1998, \apj, 496, 407 

 \bibitem[H\o g et al. (2000)]{hog00} H\o g, E., Fabricius, C., Makarov, V.V. et al., 2000, A\&A 355, L27 
 
 \bibitem[Holgado et al. (2018)]{holgado18} Holgado, G., Sim\'on-D\'iaz, S., Barb\'a, R.H. et al., 2018, \aap, in press
 
 \bibitem[Langer \& Kudritzki (2014)]{langer14} Langer, N. \& Kudritzki, R.P. \ 2014, \aap, 564, A52   
 
 \bibitem[Lindegren et al.(2018a)]{lindegren18} Lindegren, L., Hernandez, J., Bombrun, A., et al.\ 2018a, A\&A, 616, A2 

 \bibitem[Lindegren et al.(2018b)]{lindegren18b} Lindegren, L., Hernandez, J., Bombrun, A., et al.\ 2018b, in XXX IAU General Assembly, Vienna, 18-31 August 2018 

 \bibitem[Ma\'iz Apell\'aniz et al. (2014)]{maiz14} Ma{\'{\i}}z Apell{\'a}niz J., Evans, C.J., Barb{\'a}, R.H. et al., 2014, \aap, 564, A63

 \bibitem[Ma\'iz Apell\'aniz et al. (2016)]{maiz16} Ma{\'{\i}}z Apell{\'a}niz J., et al., 2016, ApJS, 224, 4 

 \bibitem[Ma\'iz Apell\'aniz \& Barb\'a (2018)]{maiz18} Ma{\'{\i}}z Apell{\'a}niz J. \& Barb\'a., R.H., 2018, \aap, 613, A9

 \bibitem[Markova \& Puls (2008)]{markova08} Markova, N. \& Puls, J. \ 2008, \aap 478, 823
 
 \bibitem[Markova et al.(2018)]{markova18} Markova, N., Puls, J., Langer, N. \ 2018, \aap, 613, A12 

 \bibitem[Martins et al.(2005)]{martins05} Martins, F., Schaerer, D., Hillier, D.~J. \ 2005, \aap, 436, 1049 
 
 \bibitem[Massey et al.(2013)]{massey13} Massey, P., Neugent, K.F., Hillier, D.J. \& Puls, J. \ 2013, \apj, 768, 6 
 
 \bibitem[Mello et al. (2012)]{mello12} Mello, D.R.C., Daflon, S., Pereira, C.B. \& Hubeny, I., 2012, A\&A 543, A11
 
 \bibitem[McCausland et al. (1992)]{McCausland92} McCausland R.~J.~H., Conlon E.~S., Dufton P.~L., Keenan F.~P., 1992, ApJ, 394, 298
 
 \bibitem[Miller Bertolami (2016)]{bertolami16} Miller Bertolami, M.M. \ 2016, \aap, 588, A25

 \bibitem[Mokiem et al.(2007)]{mokiem07} Mokiem, M.~R., de Koter, A., Vink, J.~S., et al.\ 2007b, \aap, 473, 603  
 
 \bibitem[Parthasarathy (1993a)]{partha93a} Parthasarathy,M. 1993a, \apj, 414, L109
 
 \bibitem[Parthasarathy (1993b)]{partha93b} Parthasarathy,M., 1993b, ASPC 45, 173
 
 \bibitem[Parthasarathy et al. (2000)]{partha00} Parthasarathy M., Vijapurkar J., Drilling J.~S., 2000, A\&AS, 145, 269

 \bibitem[Parthasarathy et al. (1993)]{partha93c} Parthasarathy M., Garcia-Lario P., Pottasch S.~R., Manchado A., Clavel J., de Martino D., van de Steene G.~C.~M., Sahu K.~C., 1993, A\&A, 267, L19

  \bibitem[Parthasarathy et al. (1995)]{partha95} Parthasarathy M., et al., 1995, A\&A, 300, L25 
  
 \bibitem[Prinja et al. (2012)]{prinja12} Prinja, R.K., Masa, D.L., Urbaneja, M.A. \& Kudrtizki, R.P., 2012, \mnras, 422, 3142
 
  \bibitem[Puls et al.(1996)]{puls96} Puls, J., Kudritzki, R.P., Herrero, A. et al., 1996, A\&A, 305, 171

  \bibitem[Puls et al.(2005)]{puls05} Puls J., Urbaneja M.~A., Venero R., Repolust T., Springmann U., Jokuthy A., Mokiem M.~R., 2005, A\&A, 435, 669 
  
  \bibitem[Ram\'irez-Agudelo et al.(2013)]{ramirezag13} Ram\'irez-Agudelo, O.H., Sim\'on-D\'iaz, S., Sana, H. et al. \ 2013, \aap, 560, A29
  
  \bibitem[Reid et al. (2014)]{reid14} Reid, M.J., Menten, K.M., Brunthaler, A. et al., 2014, \apj, 783, 130 

  \bibitem[Renzo et al. (2019)]{renzo19} Renzo, M., Zapartas, E., de Mink, S.E. et al., 2019, A\&A 624, A66
  
  \bibitem[Repolust et al.(2004)]{repolust04} Repolust, T., Puls, J. \& Herrero, A., 2004, A\&A, 415, 349
  
  \bibitem[Sab\'in-Sanjuli\'an et al.(2014)]{sabin14} Sab\'in-Sanjuli\'an, C., Sim\'on-D\'iaz, S., Herrero, A. et al. \ 2014, \aap, 564, A39
  
  \bibitem[Sab\'in-Sanjuli\'an et al.(2017)]{sabin17} Sab\'in-Sanjuli\'an, C., Sim\'on-D\'iaz, S., Herrero, A. et al. \ 2017, \aap, 601, A79
 
  \bibitem[Santolaya-Rey et al.(1997)]{santolaya97} Santolaya-Rey, A.E., Puls, J., Herrero, A. \ 1997, \aap, 323, 488
  
  \bibitem[Savage et al. (2001)]{savage01} Savage, B.D., Meade, M.R. \& Sembach, K.R. \ 2001, \apjs 136, 631
  
  \bibitem[Savage et al. (2017)]{savage17} Savage, B.D., Kim, T.-S., Fox, A.J. et al., \ 2017, \apjs 232, 25 
  
  \bibitem[Skrutskie et al. (2006)]{skrutskie06} Skrutskie, M.F., Cutri, R.M., Stiening, R. et al., 2006, AJ 131, 1163 
  
  \bibitem[Szczerba et al. (2007)]{szczerba07} Szczerba, R., Si\'odmiak, N., Stasi\'nska, G. \& Borkowski, J. \ 2007, \aap 469, 799 
  
  \bibitem[Sim\'on-D\'iaz (2010)]{ssimon10} Sim\'on-D\'iaz, S.,  2010, \aap, 510, A22 
    
  \bibitem[Sim{\'o}n-D{\'{\i}}az et al.(2011)]{ssimon11} Sim{\'o}n-D{\'{\i}}az, S., Castro, N., Garcia, M., Herrero, A., \& Markova, N.\ 2011a, Bulletin de la Societe Royale des Sciences de Liege, 80, 514  
 
 \bibitem[Sim\'on-D\'iaz \& Herrero(2014)]{SH14} Sim\'on-D\'iaz, S. \& Herrero, A. \ 2014, \aap, 582, 135 
 
 \bibitem[Townsend et al. (2004)]{townsend04} Townsend, R.H.D., Owocki, S.P., Howarth, I.D., 2004, MNRAS, 350, 189
 
\bibitem[Walborn(2009)]{walborn09} Walborn, N.R. \ 2009, in {\it Massive Stars: from Pop. III and GRBs to the Milky Way}, eds. M. Livio \& E. Villaver, STScI Symp. Ser. 20, 167 

\end{thebibliography}
\end{document}